\documentclass[preprint]{aastex6}
\def\degpoint{\ifmmode ^{\rm{o}}\!. \else $^{\rm{o}}\!.$\fi}

\newcommand{\Msun}{\mbox{$M_{\odot}$}}
\newcommand{\Rsun}{\mbox{$R_{\odot}$}}

\newcommand{\Rjup}{\mbox{$R_{\rm Jup}$}}

\newcommand{\Rearth}{\mbox{$R_{\oplus}$}}

\usepackage{graphicx}
\usepackage{subfigure}
\usepackage{longtable}

\begin{document}

\title{The K2-HERMES Survey. I. Planet Candidate Properties from 
K2 Campaigns 1-3}

%\author{Robert A.~Wittenmyer\altaffilmark{1,2}, Sanjib 
%Sharma\altaffilmark{3}, Jonathan Horner\altaffilmark{1,2}, Dennis 
%Stello\altaffilmark{2}, Jake Clark\altaffilmark{1}, Stephen R. 
%Kane\altaffilmark{4}, Galahs }

\author{Robert A. Wittenmyer}
\affiliation{University of Southern Queensland, Computational
Engineering and Science Research Centre, Toowoomba, Queensland 4350,
Australia}
\affiliation{Australian Centre for Astrobiology, University of New South Wales, Sydney, NSW 2052, Australia}
\author{Sanjib Sharma}
\affiliation{Sydney Institute for Astronomy, School of Physics,
  University of Sydney, NSW 2006, Australia}
\author{Dennis Stello}
\affiliation{School of Physics, University of New South Wales, Sydney, NSW 2052, Australia}
\affiliation{Stellar Astrophysics Centre, Department of Physics and Astronomy, Aarhus University, DK-8000 Aarhus C, Denmark}
\affiliation{Sydney Institute for Astronomy, School of Physics,
  University of Sydney, NSW 2006, Australia}
\author{Sven Buder}
\affiliation{Max Planck Institute  for Astronomy (MPIA), Koenigstuhl 17, D-69117 Heidelberg}
\affiliation{Fellow of the International Max Planck Research School for Astronomy \& Cosmic Physics at the University of Heidelberg}
\author{Janez Kos}
\affiliation{Sydney Institute for Astronomy, School of Physics,
  University of Sydney, NSW 2006, Australia}
\author{Martin Asplund}
\affiliation{Research School of Astronomy \& Astrophysics, Australian National University, ACT 2611, Australia}
\author{Ly Duong}
\affiliation{Research School of Astronomy \& Astrophysics, Australian National University, ACT 2611, Australia}
\author{Jane Lin}
\affiliation{Research School of Astronomy \& Astrophysics, Australian National University, ACT 2611, Australia}
\author{Karin Lind}
\affiliation{Max Planck Institute  for Astronomy (MPIA), Koenigstuhl 17, D-69117 Heidelberg}
\affiliation{Department of Physics and Astronomy, Uppsala University, Box 516, SE-751 20 Uppsala, Sweden}
\author{Melissa Ness}
\affiliation{Max Planck Institute  for Astronomy (MPIA), Koenigstuhl 17, D-69117 Heidelberg}
\author{Tomaz Zwitter}
\affiliation{Faculty of Mathematics and Physics, University of Ljubljana, Jadranska 19, 1000 Ljubljana, Slovenia}
\author{Jonathan Horner}
\affiliation{University of Southern Queensland, Computational
Engineering and Science Research Centre, Toowoomba, Queensland 4350,
Australia}
\author{Jake Clark}
\affiliation{University of Southern Queensland, Computational
Engineering and Science Research Centre, Toowoomba, Queensland 4350,
Australia}
\author{Stephen R. Kane}
\affiliation{Department of Earth Sciences, University of California, Riverside, CA 92521, USA}
\author{Daniel Huber}
\affiliation{Institute for Astronomy, University of Hawai`i, 2680 Woodlawn Drive, Honolulu, HI 96822, USA}
\affiliation{Sydney Institute for Astronomy (SIfA), School of Physics, University of Sydney, NSW 2006, Australia}
\affiliation{SETI Institute, 189 Bernardo Avenue, Mountain View, CA 94043, USA}
\affiliation{Stellar Astrophysics Centre, Department of Physics and Astronomy, Aarhus University, Ny Munkegade 120, DK-8000 Aarhus C, Denmark}
\author{Joss Bland-Hawthorn}
\affiliation{Sydney Institute for Astronomy, School of Physics, University of Sydney, NSW 2006, Australia}
\author{Andrew R. Casey}
\affiliation{Monash Centre for Astrophysics, School of Physics \& Astronomy, Monash University, Clayton 3800, Victoria, Australia}
\author{Gayandhi M. De Silva}
\affiliation{Sydney Institute for Astronomy, School of Physics, University of Sydney, NSW 2006, Australia}
\affiliation{Australian Astronomical Observatory, 105 Delhi Rd, North Ryde, NSW 2113, Australia}
\author{Valentina D'Orazi}
\affiliation{INAF Osservatorio Astronomico di Padova, vicolo dell'Osservatorio 5, 35122, Padova, Italy}
\author{Ken Freeman}
\affiliation{Research School of Astronomy \& Astrophysics, Australian National University, ACT 2611, Australia}
\author{Sarah Martell}
\affiliation{School of Physics, University of New South Wales, Sydney, NSW 2052, Australia}
\author{Jeffrey D. Simpson}
\affiliation{Australian Astronomical Observatory, 105 Delhi Rd, North Ryde, NSW 2113, Australia}
\author{Daniel B. Zucker}
\affiliation{Department of Physics \& Astronomy, Macquarie University, Sydney, NSW 2109, Australia}
\affiliation{Research Centre in Astronomy, Astrophysics \& Astrophotonics, Macquarie University, Sydney, NSW 2109, Australia}
\affiliation{Australian Astronomical Observatory, 105 Delhi Rd, North Ryde, NSW 2113, Australia}
\author{Borja Anguiano}
\affiliation{Department of Astronomy, University of Virginia, Charlottesville, VA 22904-4325, USA}
\affiliation{Department of Physics \& Astronomy, Macquarie University, Sydney, NSW 2109, Australia}
\author{Luca Casagrande}
\affiliation{Research School of Astronomy \& Astrophysics, Australian 
National University, Cotter Road, Weston Creek, ACT 2611 Australia}
\author{James Esdaile}
\affiliation{School of Physics, University of New South Wales, Sydney, NSW 2052, Australia}
\author{Marc Hon}
\affiliation{School of Physics, University of New South Wales, Sydney, NSW 2052, Australia}
\author{Michael Ireland}
\affiliation{Research School of Astronomy \& Astrophysics, 
Australian National University, Cotter Road, Weston Creek, ACT 2611 
Australia}
\author{Prajwal R. Kafle}
\affiliation{International Centre for Radio Astronomy Research (ICRAR), The University of Western Australia, 35 Stirling Highway, \\Crawley, WA 6009, Australia}
\author{Shourya Khanna}
\affiliation{Sydney Institute for Astronomy, School of Physics,
  University of Sydney, NSW 2006, Australia}
\author{J.P. Marshall}
\affiliation{Academia Sinica, Institute of Astronomy and 
Astrophysics, Taipei 10617, Taiwan}
\affiliation{School of Physics, University of New South Wales, Sydney, NSW 2052, Australia}
\author{Mohd Hafiz Mohd Saddon}
\affiliation{School of Physics, University of New South Wales, Sydney, NSW 2052, Australia}
\author{Gregor Traven}
\affiliation{Faculty of Mathematics and Physics, University of Ljubljana, Jadranska 19, 1000 Ljubljana, Slovenia}
\author{Duncan Wright}
\affiliation{School of Physics, University of New South Wales, Sydney, NSW 2052, Australia}
\affiliation{Australian Astronomical Observatory, 105 Delhi Rd, North Ryde, NSW 2113, Australia}

%\author{Elaina Hyde}
%\affiliation{Western Sydney University, Locked Bag 1797, Penrith, NSW 2751, Australia}

%\altaffiltext{1}{University of Southern Queensland, Computational 
%Engineering and Science Research Centre, Toowoomba, Queensland 4350, 
%Australia}
%\altaffiltext{2}{Australian Centre for Astrobiology, UNSW Australia, 
%Sydney 2052, Australia}
%\altaffiltext{3}{Sydney Institute for Astronomy, School of Physics, 
%University of Sydney, NSW 2006, Australia}
%\altaffiltext{4}{Department of Earth Sciences, University of California 
%Riverside, 900 University Avenue, Riverside, CA 92521, US}

\email{rob.w@usq.edu.au}

\shorttitle{K2-HERMES I}
\shortauthors{Wittenmyer et al.}

%-------------------------------------------------------------------

\begin{abstract}

\noindent Accurate and precise radius estimates of transiting exoplanets 
are critical for understanding their compositions and formation 
mechanisms. To know the planet, we must know the host star in as much 
detail as possible.  We present first results from the K2-HERMES 
project, which uses the HERMES multi-object spectrograph on the 
Anglo-Australian Telescope to obtain R$\sim$28,000 spectra of up to 360 
stars in one exposure.  This ongoing project aims to derive 
self-consistent spectroscopic parameters for about half of K2 target 
stars.  We present complete stellar parameters and isochrone-derived 
masses and radii for 46 stars hosting 57 K2 candidate planets in 
Campaigns 1-3.  Our revised host-star radii cast severe doubt on three 
candidate planets: EPIC\,201407812.01, EPIC\,203070421.01, and 
EPIC\,202843107.01, all of which now have inferred radii well in excess 
of the largest known inflated Jovian planets.

\end{abstract}


%--------------------------------------------------------------------

\section{Introduction}

A little over two decades ago, the first planets were discovered 
orbiting Sun-like stars \citep{51Peg,70Vir,47Uma}, and humanity entered 
the exoplanet era. Those first planets revolutionised our understanding 
of planet formation, and offered a tantalising hint to the uniqueness of 
the architecture of our Solar system.

In the years that followed, the number of known exoplanets grew - and 
the surprises continued to come. Some systems contained planets moving 
on highly eccentric orbits \citep[e.g.][]{ecc1,ecc2,ecc3}. Others had 
multiple planets on highly compact orbits, far closer to their host 
stars than the distance between Mercury and the Sun 
\citep[e.g.][]{Compact1,Compact2,Compact3}, or planets on highly 
inclined orbits \citep[e.g.][]{RM1,RM2,RM3}.

However, the techniques used to discover these myriad planets are all 
strongly biased - towards more massive planets, and typically towards 
planets with short orbital periods \citep[e.g.][]{otoole09,bias2,bias3}. 
To begin searching for true Solar system analogues requires either a 
search for massive long-period planets 
\citep[e.g.][]{bedell15,anal1,anal2,anal3}, or a search for small worlds 
analogous to the terrestrial planets \citep[e.g.][]{howard12, etaearth, 
minerva}.

The first real information on the frequency of planets that resemble the 
Solar system's rocky worlds came from the Kepler spacecraft 
\citep{Kep1}, which stared continuously at a single patch of the sky for 
just over four years. By observing over 100,000 stars, Kepler discovered 
an unprecedented number of planets \citep{Kep2} - including objects as 
small as the planet Mercury \citep{TinyPlanet}. From Kepler's great 
census, it has become clear that small planets are common - which might 
be the first hint that our Solar system is far from unique.

In 2014, the second of Kepler's four reaction wheels, used to orient the 
spacecraft and keep it pointing at its target field of view, failed. As 
a result, the first phase of Kepler's mission came to an end, and the 
spacecraft was repurposed to carry out the K2 survey \citep{howell14}. 
The telescope was re-oriented to point in the plane of the ecliptic - a 
position it could maintain without requiring the use of the broken 
reaction wheels.

To avoid pointing directly at the Sun, the K2 mission points at any 
given patch of the ecliptic for a period of about 80 days. At the end of 
that observing cycle, it pivots further away from the Sun, to point at a 
new field, and repeat the process. Where the original \textit{Kepler} 
survey searched for planets out to orbital periods of around one year, 
K2 can find only those planets with the shortest orbital periods. 
However, to counterbalance this, through the full K2 mission, the 
spacecraft will be able to observe a diverse multitude of 
community-selected stars, and thereby yield a new treasure trove of 
short period planets to add to the original survey's grand census 
\citep[e.g.][]{v16, c16}.

To fully understand the nature of the planets found orbiting those 
stars, it is critically important that the stars themselves are well 
characterised and understood \citep[e.g.][]{huber14, ren16, cks1}. As a 
result, there is a need for the target stars to be observed 
spectroscopically from the ground. With such a large number of stars to 
be targeted, it would be grossly inefficient to observe them one at a 
time - but fortunately, in Australia, the 2dF/HERMES instrument on the 
Anglo-Australian Telescope is ideally suited to such a survey.

Built to perform 'Galactic Archaeology' 
\citep[e.g.][]{martell17,GALAH2,kos17}, the 2dF/HERMES instrument allows 
observers to obtain high-quality and high-resolution ($R\sim$28,000) 
spectra of several hundred stars in a single observation - typically of 
around an hour's duration. Observations with HERMES allow the abundances 
of a number of species in the target stars to be determined, as well as 
enabling us to obtain relatively precise values for the stellar 
parameters (mass, radius, and age) - information critical to the 
understanding of the plethora of planets that will be found by the K2 
mission.

In this paper, we present the first observations from the K2-HERMES 
survey. In Section 2, we describe the observing setup for our survey, 
and give more detail on the 2dF/HERMES instrument. In section 3, we 
describe how the stellar parameters have been calculated from the HERMES 
spectra, before detailing the physical properties of the $K2$ planet 
candidates orbiting those stars in section 4.  Finally, in section 5, we 
present our conclusions, and discuss our plans for future work.

%--------------------------------------------------------------------

\section{Observations and Data Analysis}
\label{obs}

The observations were obtained with the 3.9m Anglo-Australian Telescope 
located at Siding Spring Observatory in Australia.  We use the High 
Efficiency and Resolution Multi-Element Spectrograph (HERMES), which can 
obtain spectra of up to 360 science targets simultaneously 
\citep{sheinis15, heijmans12, brzeski, barden10}

\subsection{Observational strategy}

\begin{figure}
%\plotone{k2c1_fov.eps}
\includegraphics[width=1.0\textwidth]{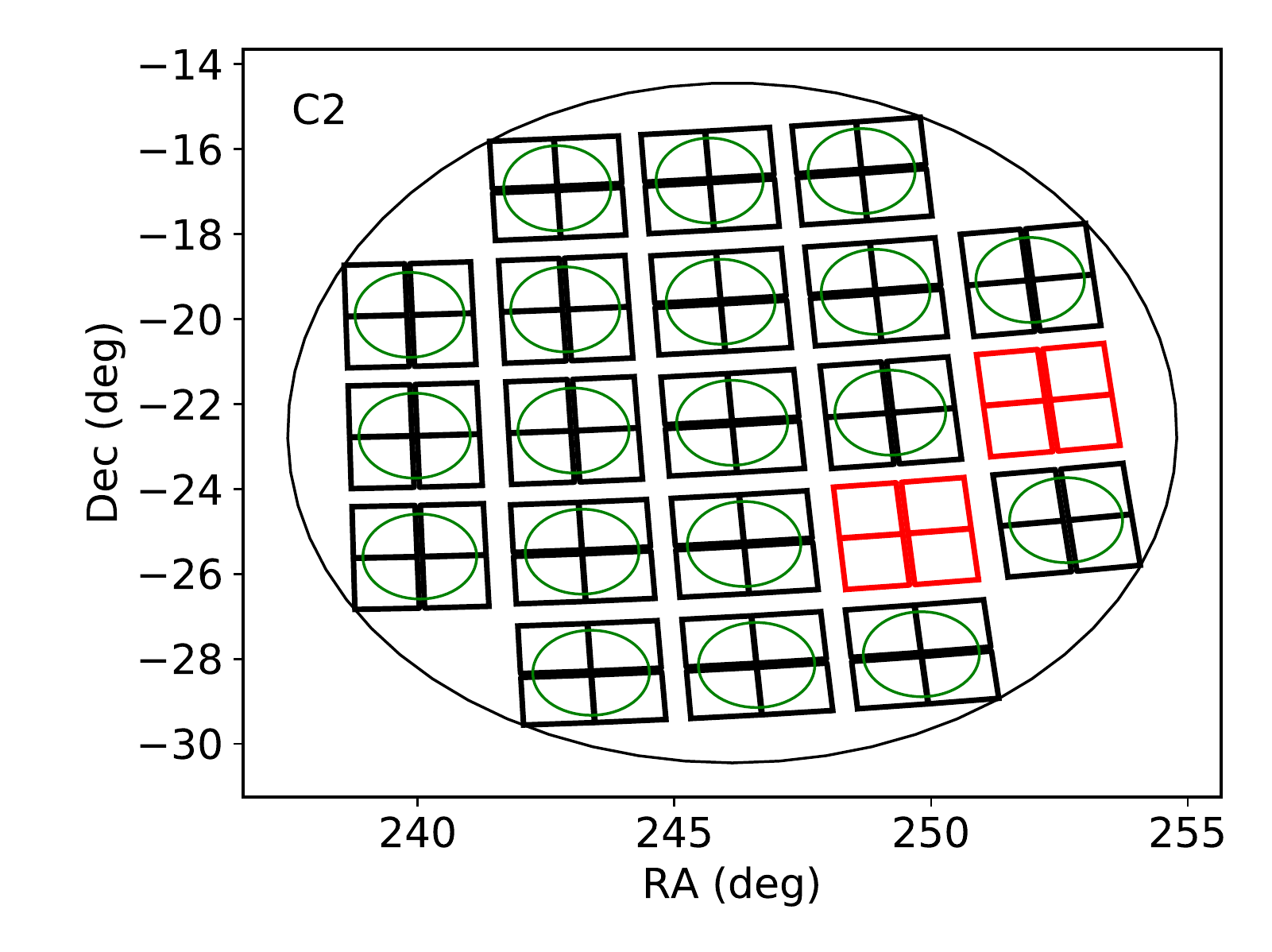}
\caption{The \textit{Kepler} field of view and the layout of its CCD 
modules, overlaid with the HERMES field of view (green circles).  The 
red modules are inoperative. }
\label{fig:fov}
\end{figure}

A single 2dF/HERMES exposure covers most of a \textit{Kepler} CCD module 
(Figure~\ref{fig:fov}), enabling relatively efficient observations of K2 
fields.  Our objective is simply to gather spectra for as many K2 
targets as possible, without introducing biases driven by the relative 
probability of hosting a planet.  As with the related TESS-HERMES 
program \citep{sharma17}, we favour bright stars to obtain targets of 
most interest for asteroseismic and exoplanetary science.  Owing to 
fibre ``cross talk'' in the instrument, we follow the procedure 
implemented by the GALAH survey \citep{martell17}, where the fields are 
chosen such that the brightest and faintest stars observed do not differ 
in brightness by more than three magnitudes.  Balancing this against the 
desire for most efficient use of the 360 HERMES science fibres results 
in a strategy whereby we observe each \textit{Kepler} CCD module twice: 
once as a bright visit ($10<V<13$) and again as a faint visit 
($13<V<15$).  Bright visits consist of a single 30-minute exposure, 
while faint visits consist of three 30-minute exposures.  In the bright 
visits, HERMES typically observes all of the available bright stars in a 
single pass.  The total number of stars per 2dF field in this range is 
typically less than 360, so we can observe all of them in one 2dF 
pointing.  In the faint visits, the total number of stars per 2dF field 
is greater than 400 but the number of $K2$ targets is 210 on average.  
This means all K2 targets lying within a 2dF circle can be observed in 
just one telescope pointing (consisting of a ``bright'' and a ``faint'' 
exposure).

\subsection{Raw reduction}

\begin{figure}
%\plotone{Signal2Noise.pdf}
\includegraphics[width=1.0\textwidth]{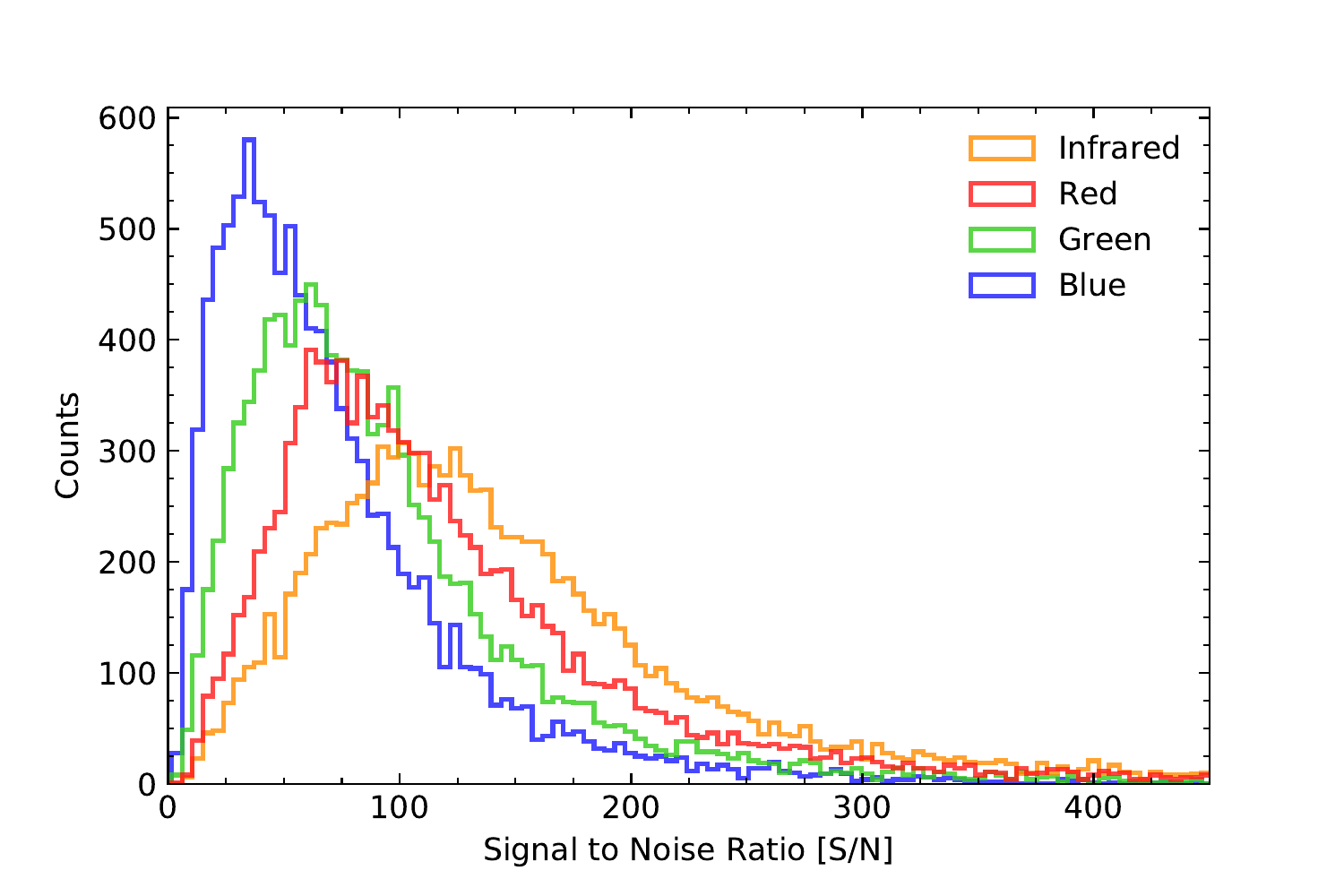}
\caption{Signal to noise per resolution element for the four HERMES 
bandpasses. }
\label{fig:sn}
\end{figure}

The K2-HERMES survey uses the same instrument as the GALAH survey 
\citep{desilva15, martell17}, and follows a similar observing strategy.  
Hence, we use the same reduction pipeline as GALAH to perform the data 
reduction from the raw CCD images to the final calibrated spectra.  The 
procedure, described fully in \citet{kos17} and \citet{sharma17}, is in 
brief: (1) raw reduction is performed with a custom IRAF-based pipeline, 
(2) four basic parameters ($T_{eff}$, log $g$, [Fe/H], and radial 
velocity) and continuum normalisation are calculated with a custom 
pipeline ``GUESS'' by matching the observed normalized spectra to 
synthetic templates.  A grid of AMBRE synthetic spectra is used for this 
purpose \citet{delaverny12}.  Figure~\ref{fig:sn} shows a histogram of 
the signal-to-noise (S/N) for our K2-HERMES spectra in each of the four 
HERMES bandpasses.

%-------------------------------------------------------------------

\section{Determination of Stellar Parameters}

The spectroscopic stellar parameters have been estimated with a 
combination of classical spectrum synthesis for a representative 
reference set of stars and a data-driven approach to propagate the 
high-fidelity parameter information with higher precision onto all the 
stars in the K2-HERMES survey.  The method is identical to that used by 
the TESS-HERMES survey \citep{sharma17}, and is briefly outlined as 
follows.  First, we use the spectrum synthesis code Spectroscopy Made 
Easy (SME) by \citet{piskunov17} to analyse the reference set.  This 
training set includes samples of stars with external parameter 
estimates, Gaia benchmark FGK stars, and stars with asteroseismic 
information from K2 Campaign 1 \citep{stello17}.  Next, we use these SME 
results as training labels of the training set as input for \textit{The 
Cannon} \citep{ness15} to propagate the analysis to all stars.  
As shown by \citet{sharma17}, the comparison with benchmark 
stars shows trends and systematic offsets for the hottest/coolest dwarfs 
and turnoff/subgiant stars of our survey, which are however below 
$250\,\mathrm{K}$ (for $T_\mathrm{eff}$) and $0.25\,\mathrm{dex}$ (for 
$\log g$).  Similar shortcomings have been noted by \citet{torres12} as 
well as \citet{bensby14}.  With new data provided by Gaia, we will 
however be able to overcome such shortcomings of purely spectroscopic 
analyses in future data releases for the joint GALAH/K2/TESS pipeline 
for HERMES.  A more detailed explanation will be given by Buder et al. 
(in prep), who include e.g. Gaia parallaxes to constrain $\log g$ for 
the training set of the GALAH Data Release 2.

\begin{figure}
\plottwo{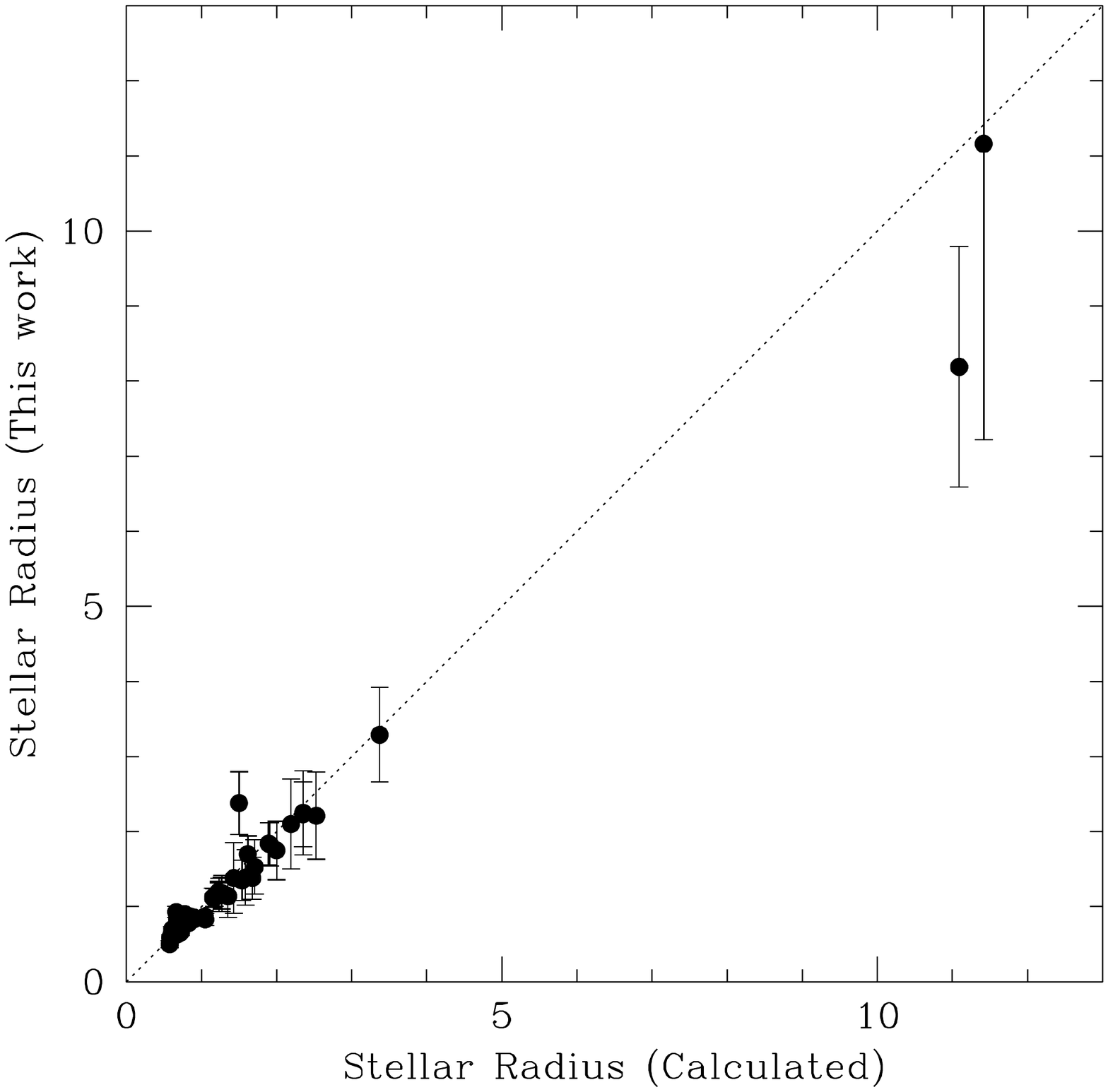}{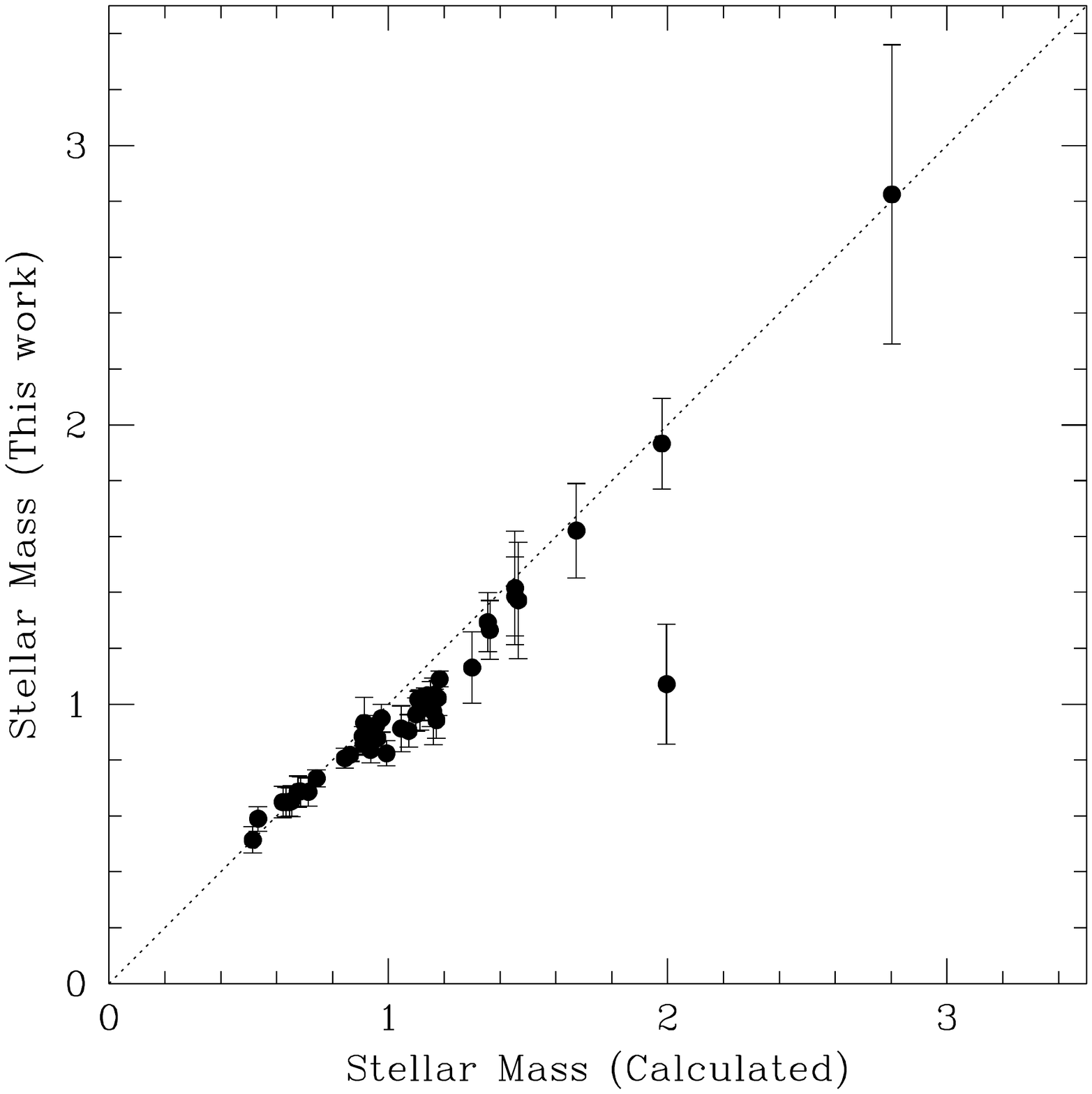}
\caption{Comparison of our derived stellar radii (left) and masses 
(right) with those estimated from the empirical relations of 
\citet{torres10}. }
\label{satisfyreferee}
\end{figure}

Derived stellar properties such as mass, radius, and age are then 
computed via the Bayesian Stellar Parameters estimator (BSTEP), 
described fully in \citet{sharma17}.  The input observables {$J$, 
$J-K_s$, $T_{eff}$, log $g$, [Fe/H]} are brought to bear on a grid of 
about $5\times10^6$ points in {[Fe/H], age, initial mass}, outputting 
the intrinsic parameters {[Fe/H], age, initial mass, distance, $E(B-V)$} 
and their probabilities.  The output is then used to compute other 
derived parameters, like stellar mass and radius, which are functions of 
the intrinsic parameters.  In Table~\ref{stellarparams} we report our 
results for the 46 planet-candidate host stars observed by K2-HERMES 
during C1-C3.  Figure~\ref{satisfyreferee} gives a comparison of 
our derived radii and masses with those obtained by the empirical method 
of \citet{torres10}.  There is good agreement within uncertainties, and 
no systematic trends are evident.

While none of the planet candidate host stars discussed here are 
\textit{a priori} known to be binaries, five are flagged as possible 
binaries from the t-SNE classification as described in \citet{GALAH2}.  
Closer inspection of the spectra reveals that four of these stars 
display evidence for a second set of lines: EPIC\,201407812, 202634963, 
202688980, and 203753577.  These stars have been marked as binaries in 
Table~\ref{stellarparams}, and the presence of a weak secondary set of 
lines may have affected \textit{The Cannon} analysis.  Hence we caution 
that those stellar parameters, while not obviously erroneous, are 
potentially unreliable.  The fifth, EPIC\,206024342, is flagged but the 
spectrum S/N is too low to visually detect any features of a second set 
of lines.

\begin{figure}
\gridline{\fig{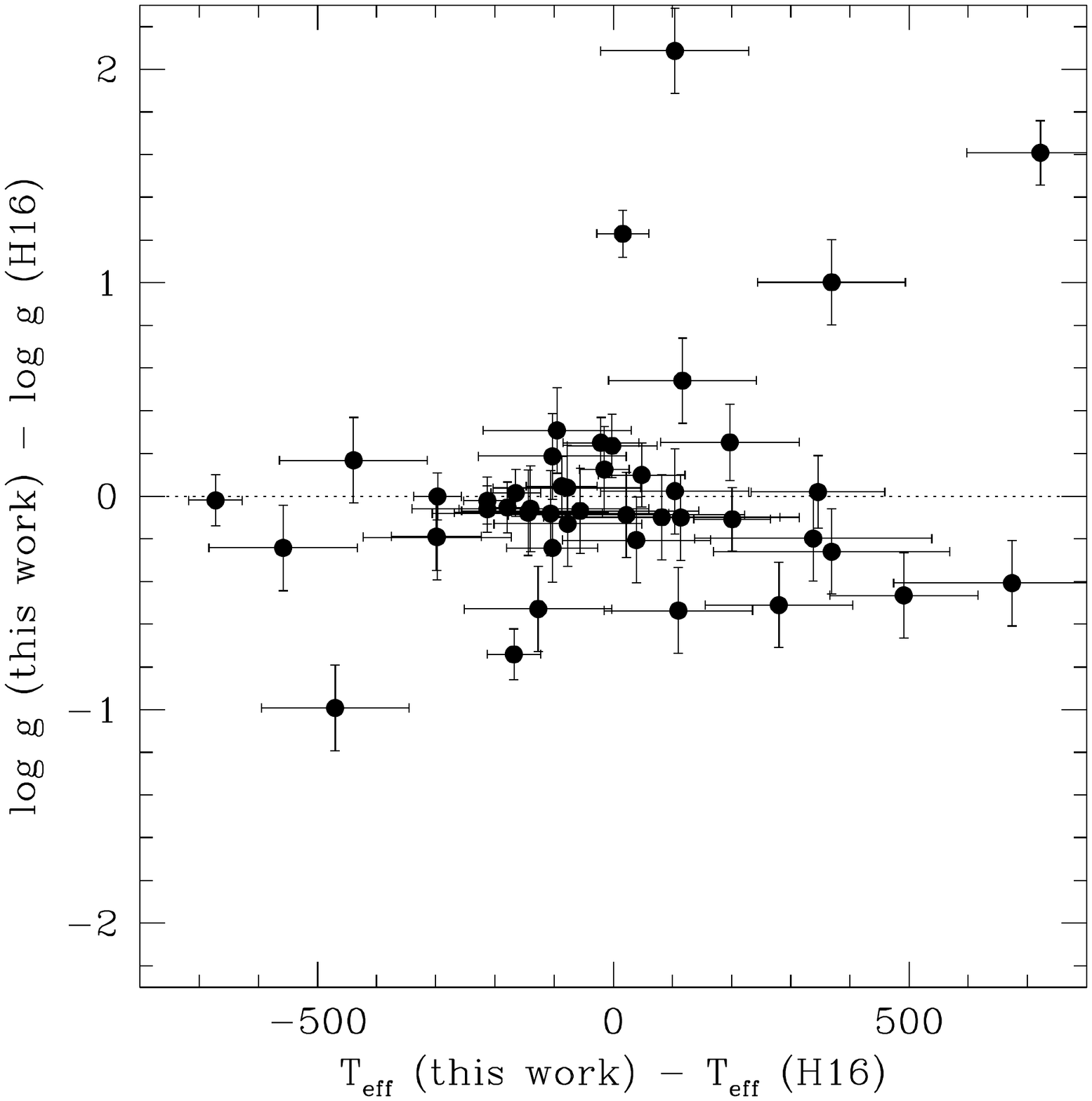}{0.33\textwidth}{(a)}
          \fig{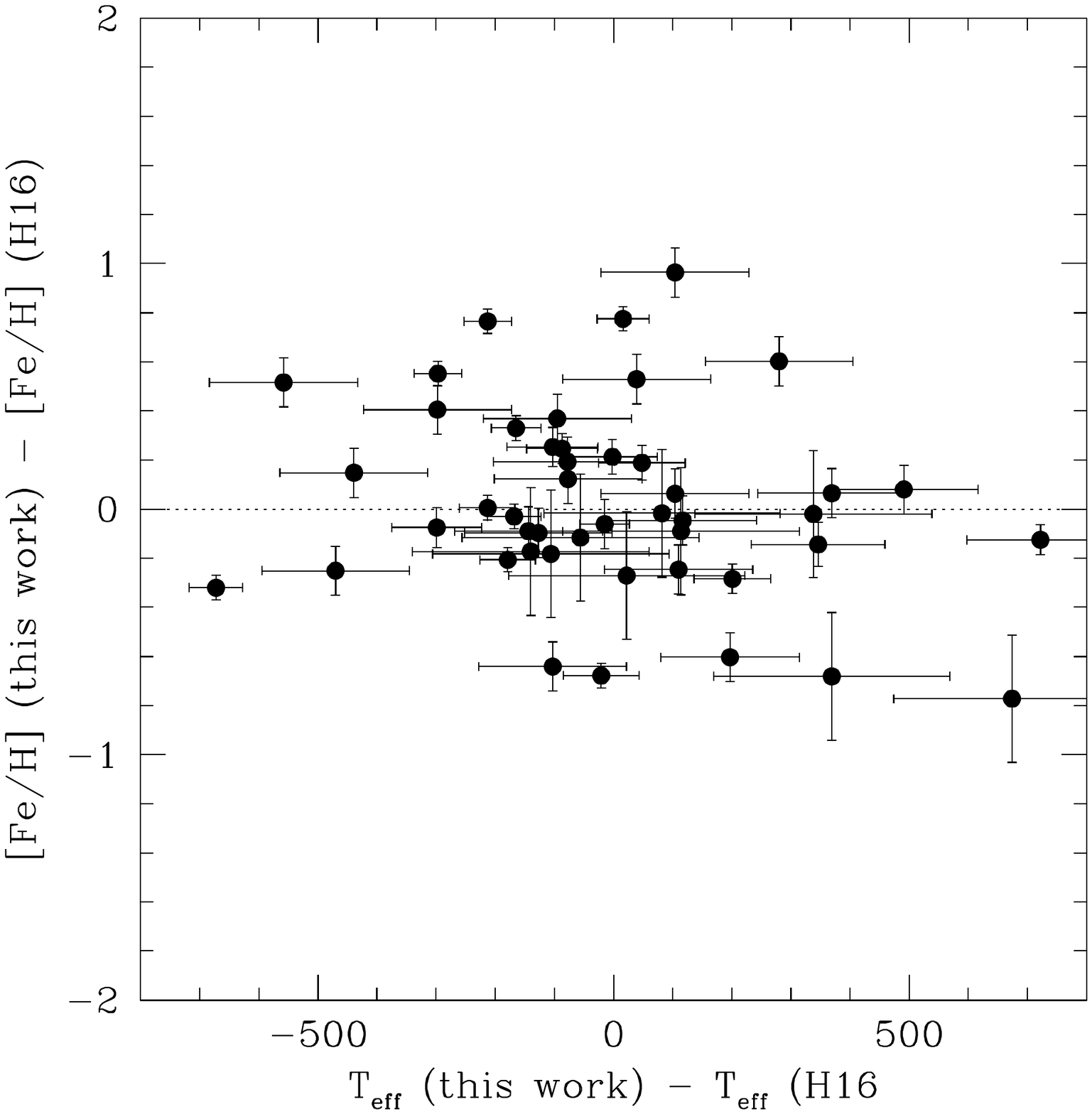}{0.33\textwidth}{(b)}
          \fig{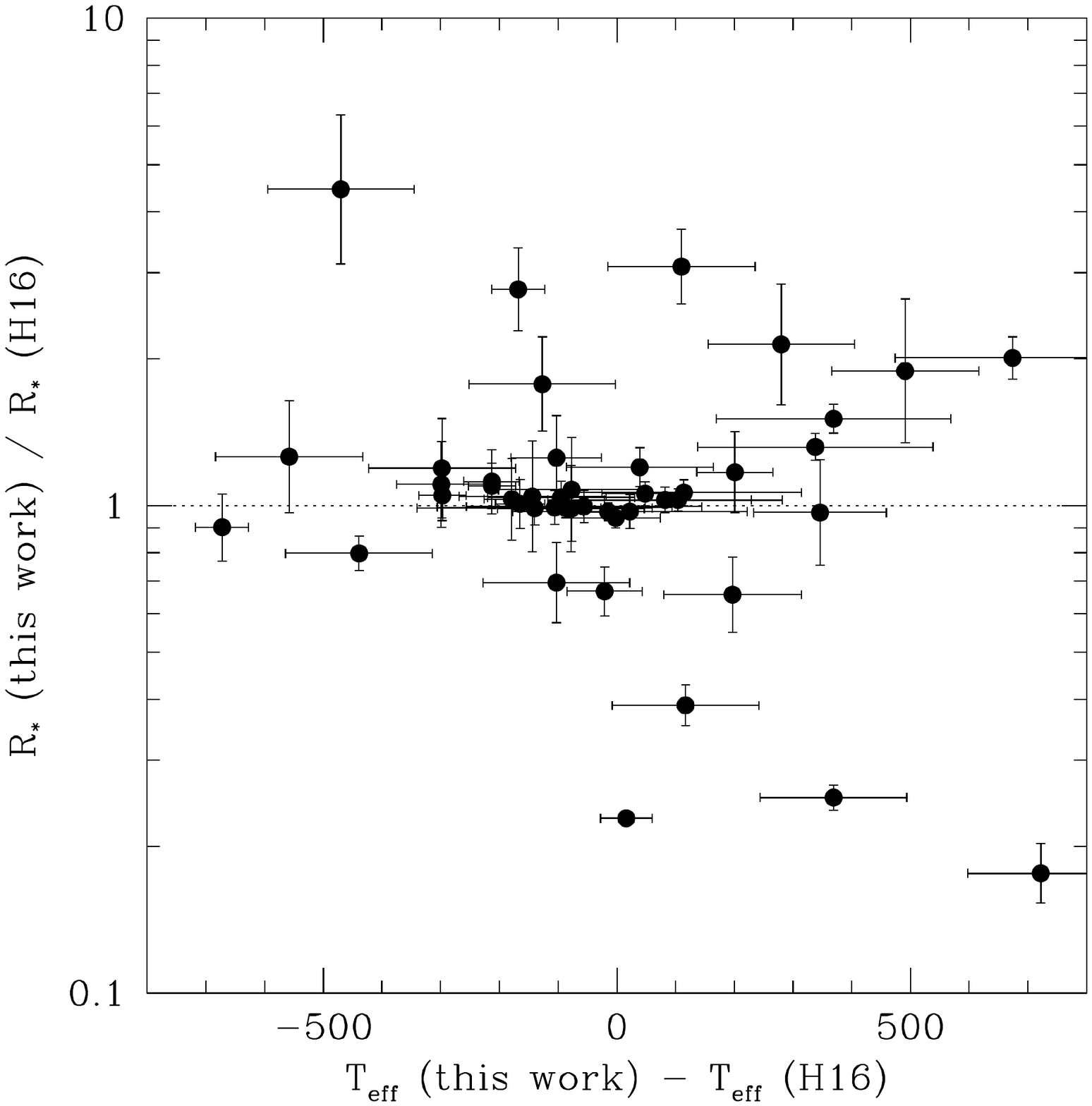}{0.33\textwidth}{(c)}}
\gridline{\fig{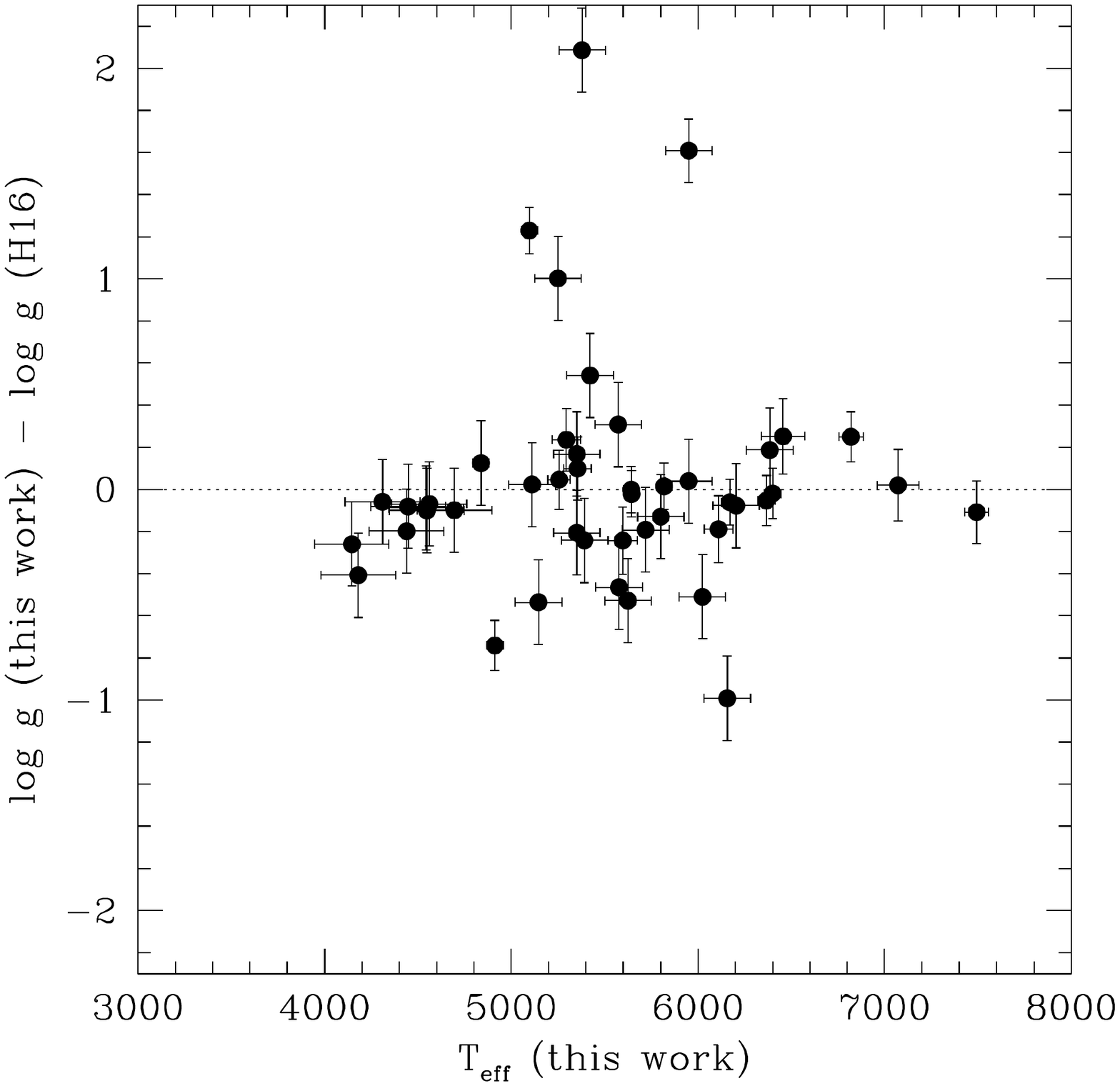}{0.33\textwidth}{(d)}
          \fig{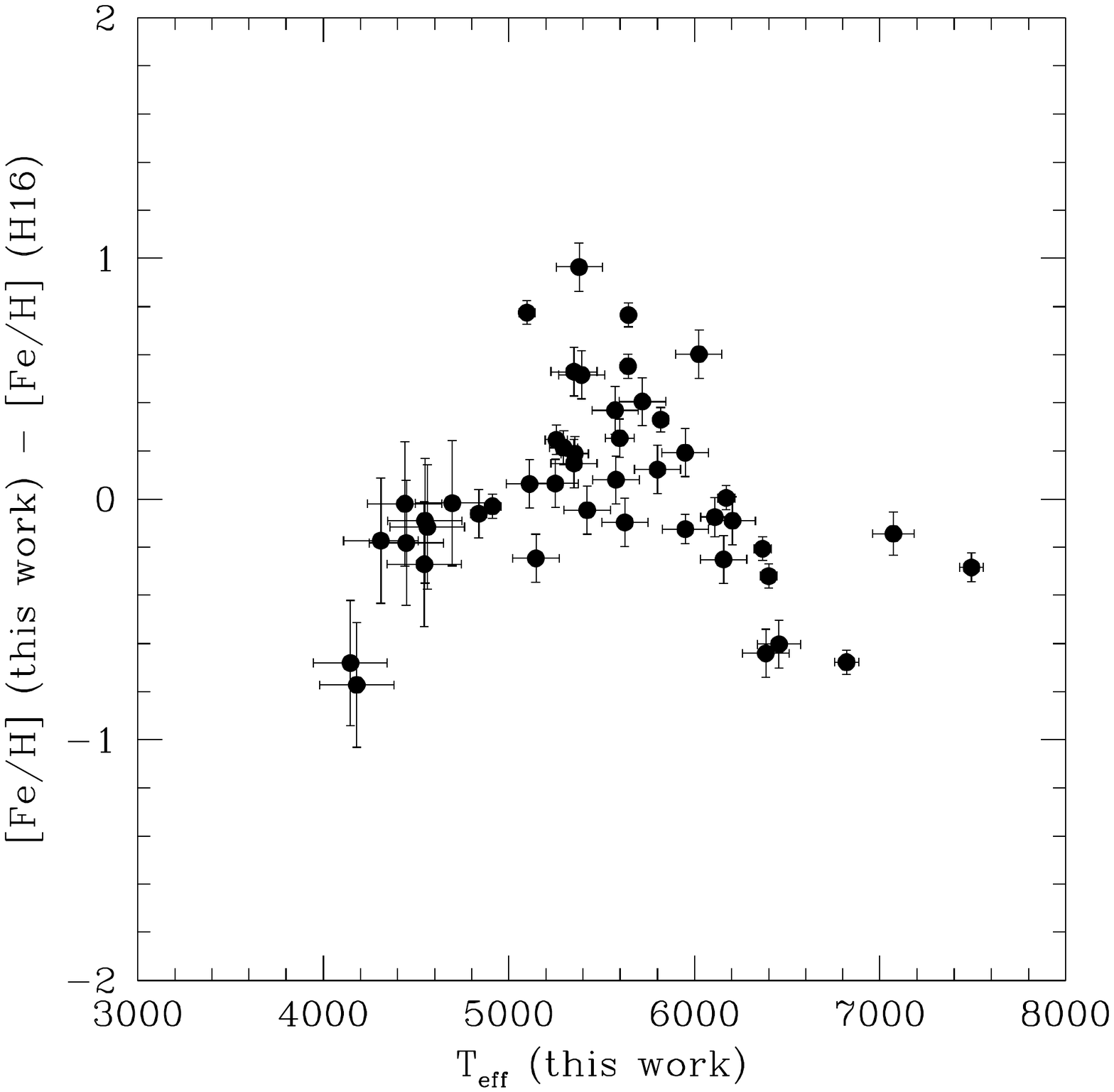}{0.33\textwidth}{(e)}
          \fig{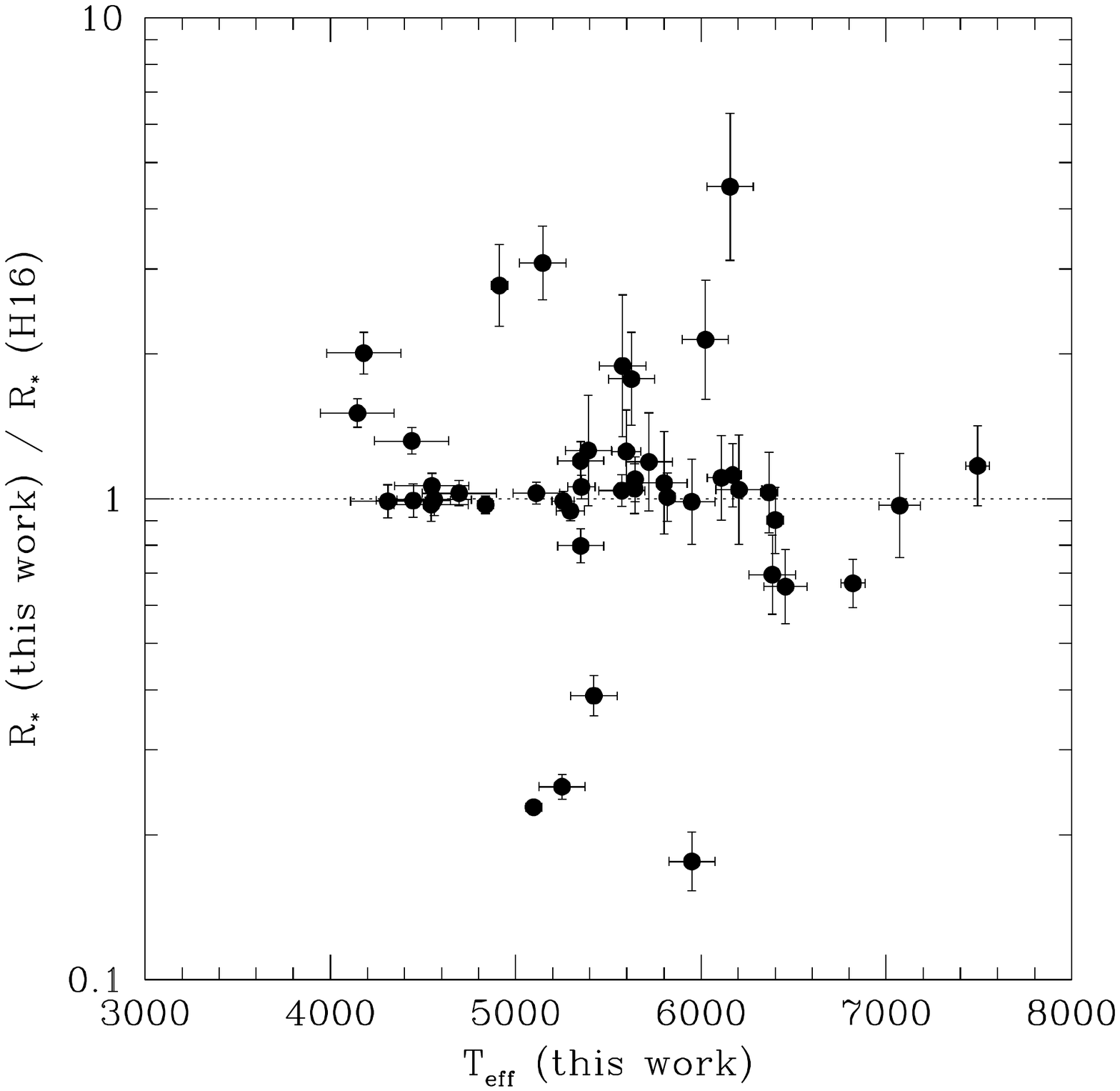}{0.33\textwidth}{(f)}}
\caption{Comparison of our spectroscopic stellar parameters with those 
of \citet{huber16}, as a function of the differences in $T_{eff}$ (top 
row) and the $T_{eff}$ derived herein (bottom row).  Error bars shown 
are those from this work.  Panels (a),(d): log $g$, panels (b),(e): 
[Fe/H], panels (c),(f): stellar radius. }
\label{stars123}
\end{figure}

\citet{huber16} presented a catalog of stellar parameters for 138,600 
stars in K2 campaigns 1-8.  For the vast majority of those stars, 
parameters were derived from photometry; here we compare those results 
with our spectroscopically-derived parameters.  Figure~\ref{stars123} 
shows the difference in log $g$ derived here with that from 
\citet{huber16}, as a function of the difference in $T_{eff}$ from the 
two works.  The centre and right panels of Figure~\ref{stars123} show 
similar comparisons, but for [Fe/H] and the derived stellar radii, 
respectively.  No systematic trends are apparent, apart from the 
expected anticorrelation between log $g$ (panel a) and the stellar 
radius (panel c). That is, stars for which we obtain a smaller log $g$ 
will have a larger derived radius.  The median parameter offsets, in the 
sense of (this work - H16), are as follows: $\Delta\,T_{eff}=-39\pm$287 
K, $\Delta$\,log $g=-0.06\pm$0.54 dex, 
$\Delta$\,[Fe/H]$=-0.025\pm$0.392, and 
$\Delta\,R_{*}=0.04\pm$2.38\,\Rsun.

The right panels of Figure~\ref{stars123} give a comparison of 
our derived stellar radii with those of H16.  Some stars do exhibit 
significant differences in derived radii, which is mainly due to the 
difference in log $g$.  \citet{huber16} measured log $g$ from photometry 
and proper motions, introducing substantial uncertainty (as evidenced by 
the large error bars in Figure~\ref{stars123}).  We find four stars with 
{log $g$ (H16) - log $g$ (this work)}$<-0.7$.  These are all red giants 
in H16 but we classify them as dwarfs.  We also find two stars with 
large differences in the opposite direction: {log $g$ (H16) - log $g$ 
(this work)}$>0.7$.  They are EPIC\,201516974 and EPIC\,203070421.  The 
first was classified as a low-luminosity giant in H16, but our results 
place it in the red clump region with log $g=2.66\pm$0.12.  The second 
was a hot dwarf but now sits in an odd position in the (log $g$, 
$T_{eff}$) plane.  In our spectroscopic pipeline it is flagged as being 
too far away from the training set, and so we caution the reader that 
the results for that star may be suspect.

% This results in the log $g$ of subgiants being overestimated by the 
% method of H16.

%-------------------------------------------------------------------

\section{Planet Candidate Parameters}

\begin{figure}
\gridline{\fig{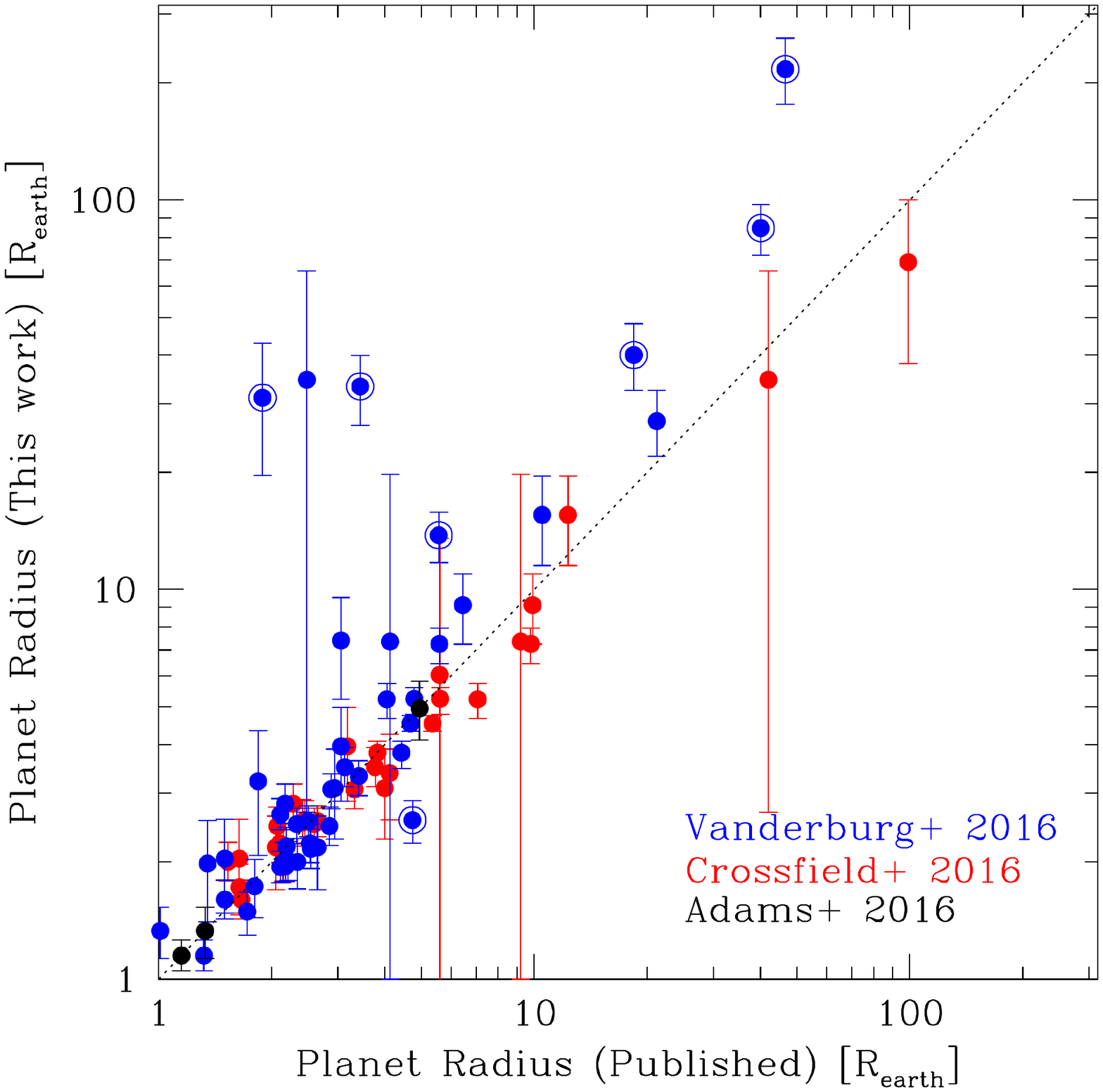}{0.5\textwidth}{(a)}
          \fig{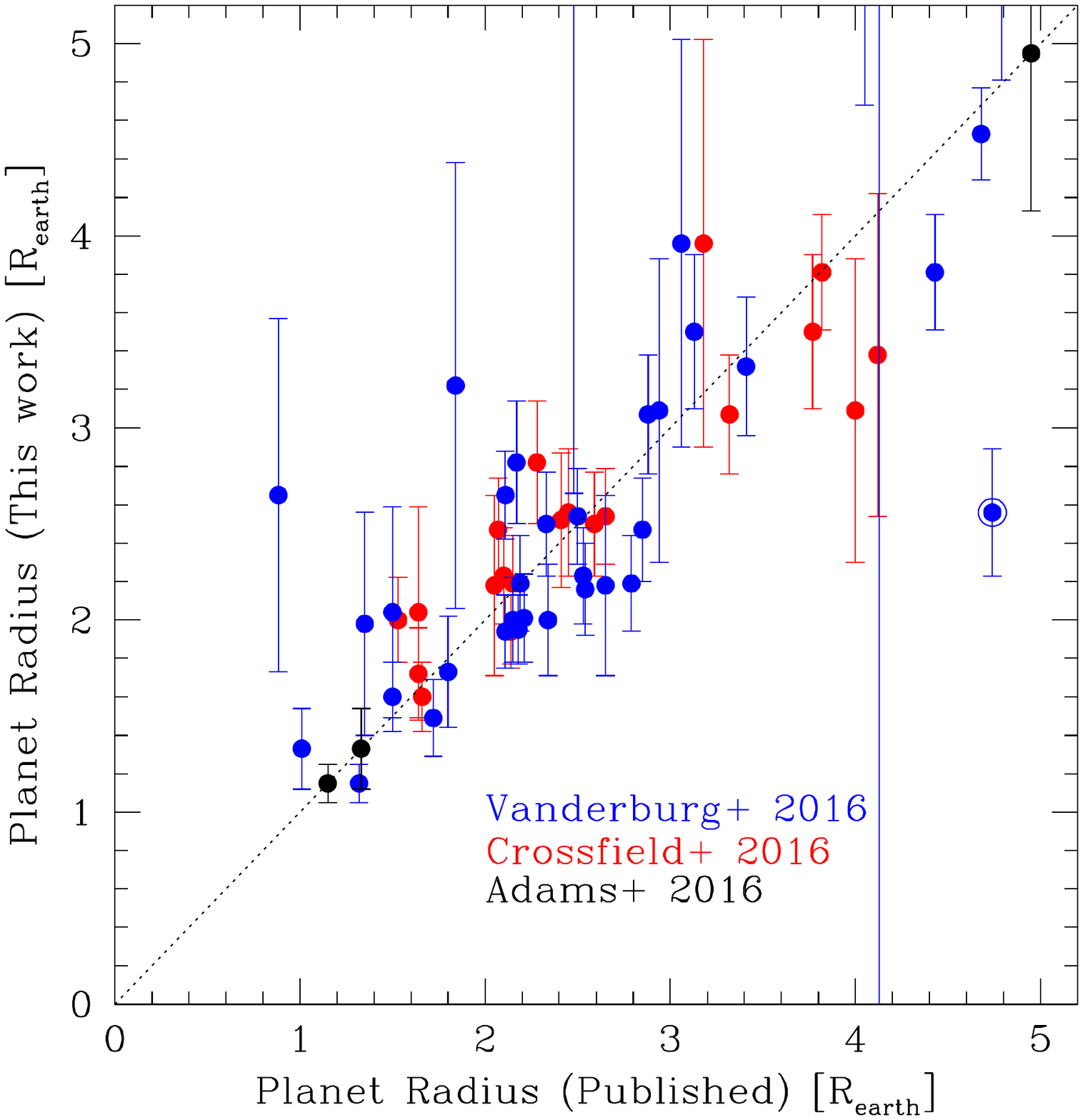}{0.5\textwidth}{(b)}}
\caption{Left panel: Radii of K2 planet candidates from C1-C3, as 
derived in this work using K2-HERMES spectra, compared with the radii as 
reported in their discovery works.  Blue: \citet{v16} -- Red: 
\citet{c16} -- Black: \citet{adams16}.  Error bars on previously 
published values are omitted for clarity.  Right panel: Same, but for 
planets smaller than 5\Rearth.  For the majority of planets, our results 
agree with the published radii, though we now find six planets with 
radii more than 3$\sigma$ larger than their published values, all from 
the \citet{v16} catalog.  Planets differing from their published 
values by more than $3\sigma$ are shown as encircled points. }
\label{fig:planets1}
\end{figure}

Table~\ref{tab:planets} gives the properties of the 57 planet candidates 
from C1-C3 for which the K2-HERMES program has obtained spectra of their 
host stars.  The planet data (orbital period and $R_p/R_*$) have been 
obtained from the NASA Exoplanet Archive.  We derived the planetary 
radii by multiplying $R_p/R_*$ by the stellar radii obtained by our 
grid-based modelling as described above.  Uncertainties in the planetary 
radii result from the propagated uncertainties in $R_*$ and $R_p/R_*$; 
for those planet candidates from \citet{v16} without published 
uncertainties in $R_p/R_*$, we adopted the median fractional uncertainty 
of 0.0025 obtained by \citet{c16}.  Figure~\ref{fig:planets1} shows our 
newly-derived planet radii against published values 
(Table~\ref{tab:planets}).

For the majority of planets, our newly-derived radii agree with the 
published values, though we do find that for six planet candidates, our 
results show radii that are more than $>3\sigma$ larger than the 
published values.  Of these, five orbit somewhat evolved stars with 
radii in the range 1.9-8\,\Rsun, resulting in larger inferred planetary 
radii, turning some potentially terrestrial worlds into gas giants.  The 
most dramatic change is candidate super-Earth EPIC\,203070421.01 
\citep{v16}, now approaching 3 Jupiter radii.  Given that the largest 
inflated planets are $\sim$2\,\Rjup, our revised host star parameters 
suggest that this candidate is a false positive.

\begin{figure}
\gridline{\fig{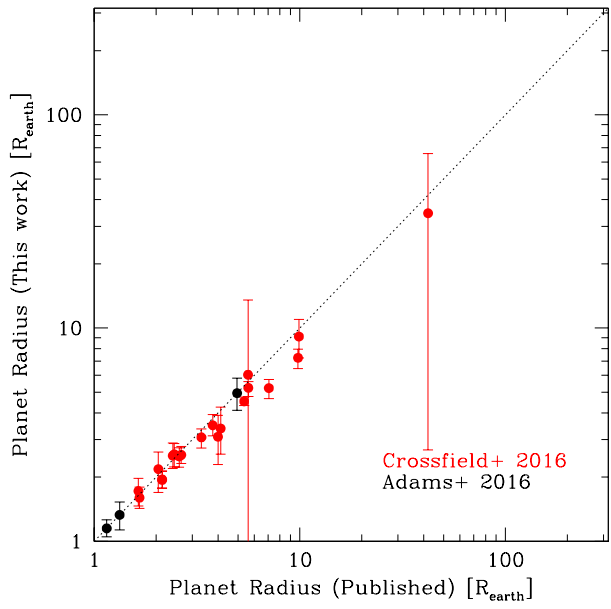}{0.5\textwidth}{(a)}
          \fig{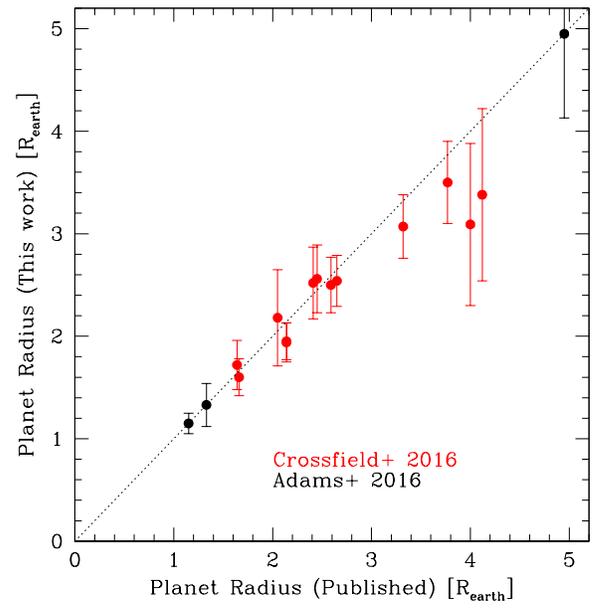}{0.5\textwidth}{(b)}}
\caption{Same as Figure~\ref{fig:planets1}, but only showing those 
planets for which the previously published radii were derived using 
spectroscopic measurements of their host stars. }
\label{newfig}
\end{figure}

The various catalogs of K2 planet candidates contain some 
fraction of stars with spectroscopically derived parameters and others 
with photometrically derived parameters only.  In Figure~\ref{newfig}, 
we reprise Figure~\ref{fig:planets1} comparing our results with 
published values, but showing only those for which the published values 
were derived from spectroscopic measurements.  Notably, all the 
candidates identified in \citet{v16} had only photometric host-star 
radii, and that catalog was the source of all of the discrepant results 
discussed above.  The remaining planets shown in Figure~\ref{newfig} are 
in excellent agreement with the published spectroscopic results.

As noted in Section 3, four stars in this sample are found to be 
binaries.  We thus caution that the following (as-yet unconfirmed) 
planet candidates may be false positives: EPIC\,201407812, 202634963, 
202688980, and 203753577.

Most of the 57 planet candidates examined here remain within the range 
of reasonable planet radii (i.e. smaller than a few tens of Earth 
radii).  However, EPIC\,202843107.01 now has a radius of 216.6\,\Rearth\ 
(approximately 2 solar radii).  That candidate has an exceptionally deep 
transit ($R_p/R_*=0.6032$), again unphysically large for a planet, 
particularly given that the host star appears to be a main-sequence A 
star (Table~\ref{stellarparams}).  We thus strongly suspect 
EPIC\,202843107.01 is a false positive.  Similarly, 
EPIC\,201407812 hosts a candidate 84.7\,\Rearth\ planet, (nearly 8 
Jupiter radii); given that the host is now confirmed as a binary, this 
candidate also appears to be a false positive. 

Our revised stellar parameters bring to light some interesting 
individual planets in this sample.  One metal-poor star hosts a 
candidate giant planet: EPIC\,206311743 ([Fe/H]$=-0.42\pm$0.10).  A 
second star, EPIC\,202634963, also hosts a candidate planet but we 
confirm it to be a double-lined spectroscopic binary, and hence is 
likely to be a false positive.  Such planets are rare by virtue of the 
well-known planet-metallicity correlation \citep{gon97,fv05}, whereby 
giant planets have difficulty forming by core accretion from metal-poor 
protoplanetary disks.  While both remain candidates, if they were to be 
confirmed, they would be extremely interesting counterexamples.

%   EPIC\,202634963 ([Fe/H]$=-0.59\pm$0.10)

Close-in planets orbiting evolved stars are also known to be rare, with 
only 12 known within 0.5\,au from the various radial velocity surveys of 
so-called ``retired A stars'' \citep[e.g.][]{bowler10, jones15, 121056}.  
Our sample contains two giants (log $g<3.5$) each hosting one planet 
candidate.  They are EPIC\,201516974 and EPIC\,203070421, neither of 
which is flagged as a binary star.  Caution is warranted, however, since 
giant stars are intrinsically more noisy (due to granulation), and hence 
the false-positive rate of detecting transit signals is higher 
\citep{sliski14}.

% a transit planet candidate with a giant-star 
% host is highly likely to be a false positive (i.e. a blend with a 
% background eclipsing binary).

% Transit durations if it really is a planet around a giant star?

\begin{figure}
%\plotone{radius_histogram.pdf}
\includegraphics[width=1.0\textwidth]{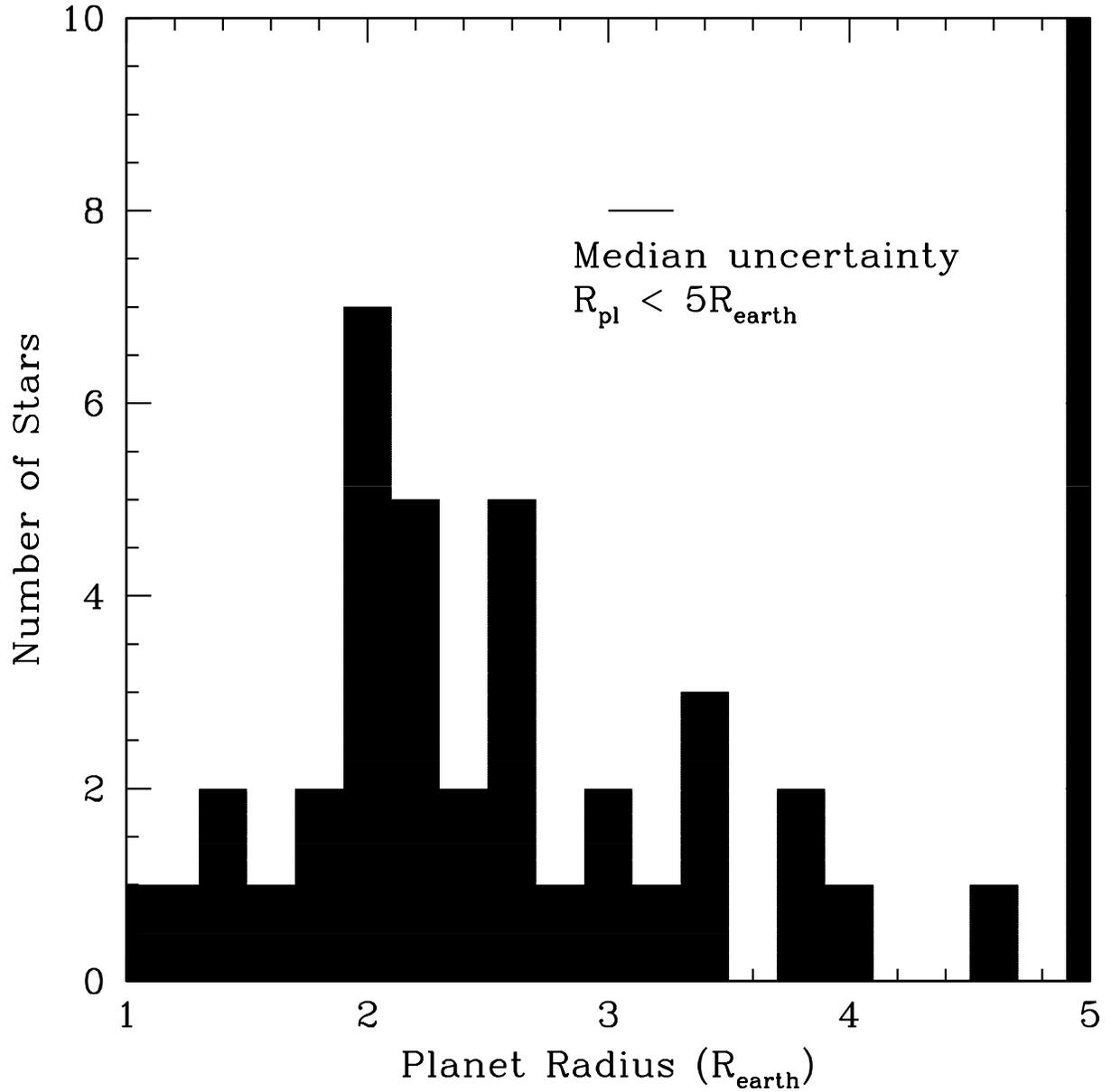}
%\gridline{\fig{radius_histogram.pdf}{0.5\textwidth}{(a)}
%          \fig{PERvsRAD.pdf}{0.5\textwidth}{(b)}}
\caption{Histogram of our derived radii for 38 K2 planet candidates 
smaller than 5\Rearth.  The median uncertainty in radius for these 
planets is shown as a horizontal bar (0.27\,\Rearth).  This distribution 
peaks at $\sim$2.0\,\Rearth\ with a secondary peak near 2.6\,\Rearth. }
\label{fig:planets2}
\end{figure}

\begin{figure}
%\plotone{PERvsRAD.pdf}
\includegraphics[width=1.0\textwidth]{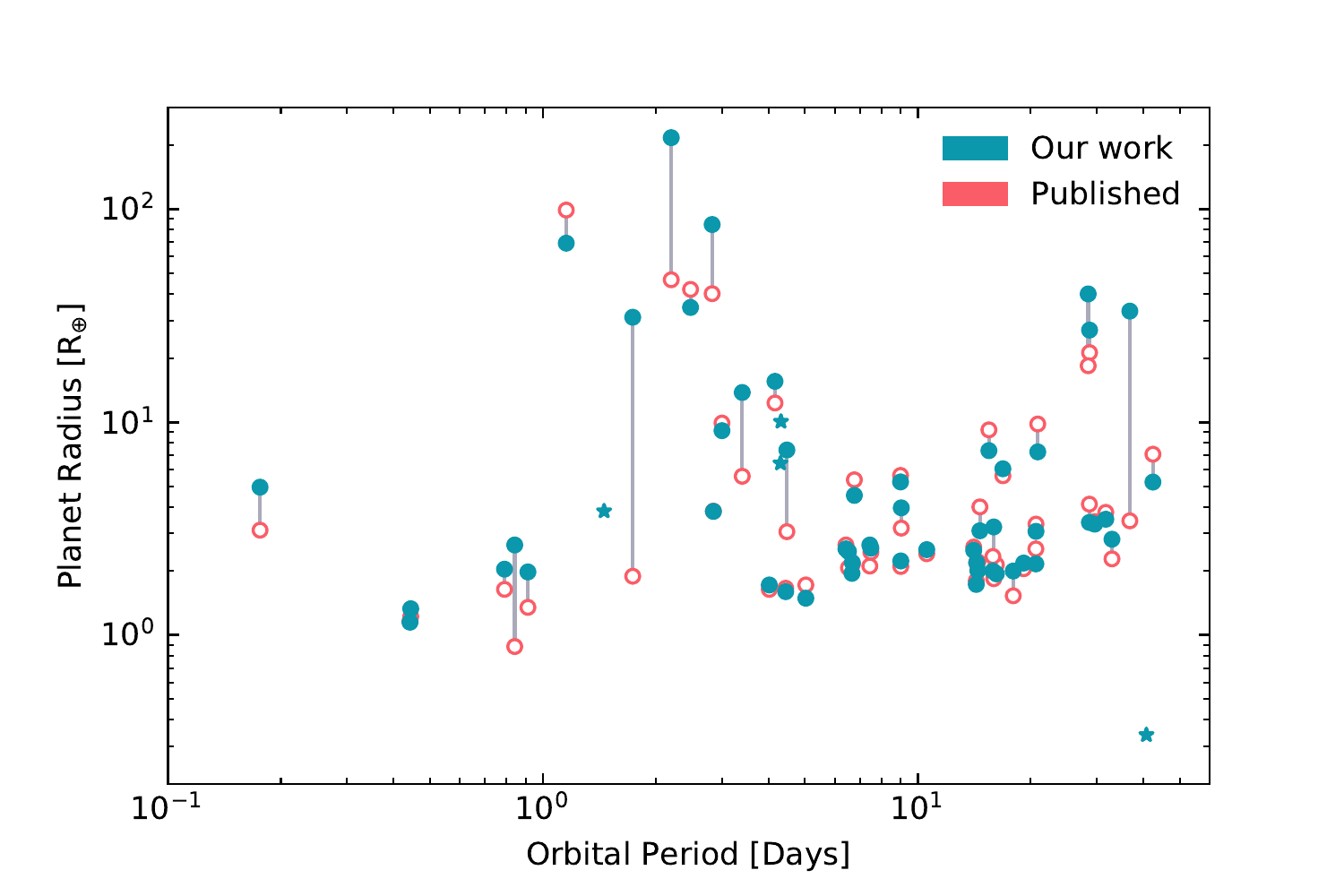}
\caption{Planet radius versus orbital period for the 57 planets examined 
here. Filled circles indicate our newly-derived radii, connected to open 
circles denoting the previously-published radii. }
\label{fig:planets3}
\end{figure}

The California-Kepler Survey (CKS) team have noted that \textit{Kepler} 
planets exhibit a gap in their radius distribution \citep{fulton17}, with 
planets of 1.5-2.0\Rearth\ apparently depleted by more than a factor of 
two.  In light of this finding, we examine the radius distribution for 
the 38 small ($R_p<5$\Rearth) $K2$ planets featured herein 
(Figure~\ref{fig:planets2}).  As perhaps expected with the low numbers 
involved, no compelling pattern is yet evident; future papers extending 
this work to further $K2$ campaigns will provide the necessary data to 
fill in this distribution.  Figure~\ref{fig:planets3} shows the planetary 
radii versus orbital period, with lines connecting the updated radii to 
their previously-published counterparts.

\begin{figure}
%\plotone{radius_vs_flux.pdf}
\includegraphics[width=1.0\textwidth]{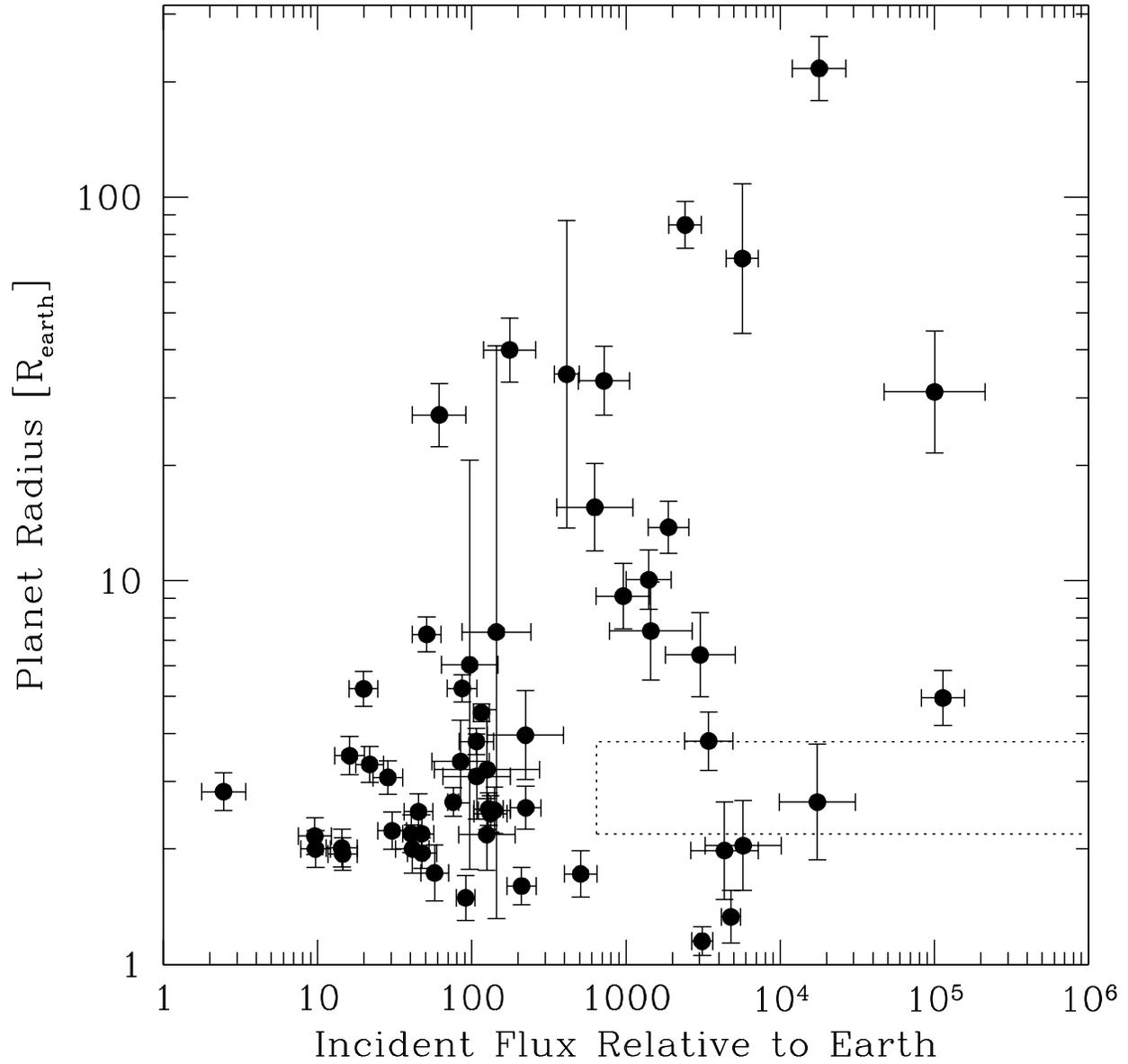}
\caption{Planet radii versus incident flux for our 57 planets.  The 
dashed lines enclose the ``hot super-earth desert'' postulated by 
\citet{lundkvist16}, denoting planets receiving more than 650 times 
Earth's incident flux, and radii from 2.2 to 3.8\,\Rearth.  We find only 
one planet falling in this region: EPIC\,203518244b. }
\label{fig:flux}
\end{figure}

The incident flux levels and equilibrium temperatures of the planet 
candidates are also shown in Table~\ref{tab:planets}.  For completeness, 
we give two values of equilibrium temperature: ``hot dayside'' and 
``well-mixed,'' corresponding to re-radiation over $2\pi$ and $4\pi$ 
steradians, respectively.  The former would be most suitable for 
close-in planets presumed to be tidally locked.  Figure~\ref{fig:flux} 
shows the relation of the planet radii and their incident fluxes for the 
57 planet candidates considered here.  \citet{lundkvist16} identified a 
``hot super-Earth desert'': a lack of planets between 2.2-3.8\,\Rearth\ 
receiving incident flux more than 650 times that of Earth.  The 
envelopes of these planets have been stripped by photoevaporation.  From 
the \textit{Kepler} planet sample for which the host stars have been 
characterised by asteroseismology, \citet{lundkvist16} found no planets 
in that range; in our sample of K2 planet candidates, we find one object 
falling within the dashed rectangle in Figure~\ref{fig:flux}.  That 
candidate, EPIC\,203518244b, orbits a star we characterise as a slightly 
evolved F subgiant ($T_{eff}=6205\pm$125\,K, log $g=3.8\pm$0.2, and 
$M_*=1.37\pm$0.21\,\Msun).  Our parameters for this star are in 
excellent agreement with those of \citet{huber16}, who derived 
$T_{eff}=6349$\,K, log $g=3.9$, and $M_*=1.42$\,\Msun.  

Though none of the planet candidates discussed here are even remotely 
considered habitable, for completeness we show the stellar habitable 
zone boundaries in Table~\ref{habzones}, which may prove relevant for 
any outer planets subsequently found to reside in these systems.

%--------------------------------------------------------------------

\section{Summary and Conclusion}

We note that $K2$ Campaigns 1-3 have more than 100 identified planet 
host stars \citep{v16, c16, adams16}, whereas this work currently 
presents spectroscopically-derived stellar properties for 57 planets 
orbiting 46 stars.  The ``missed'' planet hosts can be accounted for 
primarily by (1) targets falling on the corners of the \textit{Kepler} 
CCD modules and hence not currently covered by the K2-HERMES survey, and 
(2) host stars fainter than the faint limit of the K2-HERMES survey.  
The 2dF field is two degrees in diameter, for an area of 3.14 square 
degrees, while a single CCD module has an area of 5 square degrees.  
This means that K2-HERMES only observes $\sim$54\% of (assumed uniformly 
distributed) stars falling on $K2$ detectors.  $K2$ targets are skewed 
in favour of cooler, fainter stars \citep{huber16}, but our observed 
sample contains very few stars cooler than $\sim$4300\,K, and our 
spectroscopic analysis pipeline is currently not equipped to handle very 
cool stars.  Hence, we expect to miss out the coolest targets, 
accounting for about 25\% of the total planet candidates \citep{c16}.  
Those stars are most likely to be characterised by surveys using 
near-infrared spectroscopy \citep[e.g.][]{muirhead14, dressing1, 
dressing2}.

% we lose about 15% by faint limit alone
% lose a total of about 25% due to faint limit and colours/Teff
% since we get almost nothing cooler than ~4300 
%  Which leaves us about 47 stars from the Crossfield catalog.
%  we got 46 stars ---  notbad.jpg

The precision of our planetary radii is comparable to the published 
values.  The CKS results permitted a significant improvement in the 
precision of radii for \textit{Kepler} prime mission KOIs \citep{cks2}, 
primarily because those objects, as a sample, had very few prior 
spectroscopic observations.  On the other hand, the K2 planet candidates 
described here are small enough in number, and their host stars 
sufficiently bright, that the discovery teams have obtained spectra of 
sufficient resolution and S/N to derive reasonably precise stellar (and 
hence planetary) radii.  The primary value of our work is that we have 
presented a \textit{fully self-consistent} set of spectroscopic and 
model-derived parameters for the K2 planet candidate sample.  Applying 
this uniformly-derived set of host-star parameters to the planet 
candidate sample thus yields a set of self-consistent planetary 
properties, which is critically important for informing population 
studies of small exoplanets.  Future papers in this series will extend 
our analysis to additional K2 campaigns.  The K2-HERMES first data 
release paper (Sharma et al., in prep) will provide precise stellar 
parameters for thousands of K2 targets, including those without planet 
detections, facilitating studies of occurrence rates and planetary 
properties as functions of host-star properties and their relative 
position in the Galaxy.

%--------------------------------------------------------------------

\acknowledgements

DS is supported by Australian Research Council Future Fellowship 
FT1400147.  SS is funded by University of Sydney Senior Fellowship made 
possible by the office of the Deputy Vice Chancellor of Research, and 
partial funding from Bland-Hawthorn's Laureate Fellowship from the 
Australian Research Council.  SLM acknowledges support from the 
Australian Research Council through DECRA fellowship DE140100598.  LC is 
supported by Australian Research Council Future Fellowship FT160100402.  
This research has made use of NASA's Astrophysics Data System (ADS), and 
the SIMBAD database, operated at CDS, Strasbourg, France.  This research 
has also made use of the Exoplanet Orbit Database and the Exoplanet Data 
Explorer at exoplanets.org \citep{wright11, han14}.  We thank the 
Australian Time Allocation Committee for their generous allocation of 
AAT time which made this work possible.

\software{SME \citep{piskunov17}, The Cannon \citep{ness15} }

%--------------------------------------------------------------------

%----------------------------------------------------------

%-------------------------------------------------------------------

\clearpage

%-------------------------------------------------------------------
%  table updated 28 Sep, Cannon 1.3.4c

\begin{deluxetable}{llllll}
\tabletypesize{\scriptsize}
\tablecolumns{6}
\tablewidth{0pt}
\tablecaption{Parameters for Planet Candidate Host Stars}
\tablehead{
\colhead{EPIC ID} & \colhead{$T_{eff}$} & \colhead{log $g$} & 
\colhead{[Fe/H]} & \colhead{Mass (\Msun)} & \colhead{Radius (\Rsun)}
 }
\startdata
\label{stellarparams}
201155177 & 4695$\pm$200 & 4.56$\pm$0.20 & -0.17$\pm$0.26 & 0.69$\pm$0.05 & 0.66$\pm$0.04 \\
201291843 & 4146$\pm$200 & 4.62$\pm$0.20 & -0.47$\pm$0.26 & 0.59$\pm$0.04 & 0.59$\pm$0.04 \\
201393098 & 5625$\pm$125 & 3.90$\pm$0.20 & -0.25$\pm$0.10 & 0.97$\pm$0.12 & 1.75$\pm$0.39 \\
201403446 & 6110$\pm$76 & 4.06$\pm$0.16 & -0.42$\pm$0.08 & 0.94$\pm$0.07 & 1.38$\pm$0.28 \\
201407812\tablenotemark{a} & 5951$\pm$125 & 4.03$\pm$0.15 & -1.02$\pm$0.06 & 0.82$\pm$0.05 & 1.70$\pm$0.24 \\ %metal poor
201516974 & 4912$\pm$45 & 2.66$\pm$0.12 & -0.60$\pm$0.05 & 1.07$\pm$0.21 & 8.19$\pm$1.60 \\ %giant
201546283 & 5256$\pm$60 & 4.54$\pm$0.14 & +0.22$\pm$0.06 & 0.90$\pm$0.03 & 0.87$\pm$0.04 \\
201606542 & 5355$\pm$73 & 4.63$\pm$0.15 & +0.24$\pm$0.07 & 0.93$\pm$0.03 & 0.90$\pm$0.05 \\
% 201754305 & 4731$\pm$167 & 4.8$\pm$0.2 & -0.15$\pm$0.06 & 0.72$\pm$0.03 & 0.66$\pm$0.03 \\
201855371 & 4440$\pm$200 & 4.60$\pm$0.20 & -0.17$\pm$0.26 & 0.65$\pm$0.05 & 0.63$\pm$0.04 \\
201912552 & 4180$\pm$200 & 4.62$\pm$0.20 & -0.69$\pm$0.26 & 0.51$\pm$0.05 & 0.50$\pm$0.05 \\
202634963\tablenotemark{a} & 6385$\pm$125 & 4.30$\pm$0.20 & -0.59$\pm$0.10 & 0.96$\pm$0.06 & 1.16$\pm$0.22 \\ %metal poor with giant planet
202675839 & 5719$\pm$125 & 4.07$\pm$0.20 & +0.46$\pm$0.10 & 1.13$\pm$0.13 & 1.53$\pm$0.36 \\ %metal rich
202688980\tablenotemark{a} & 6456$\pm$117 & 4.29$\pm$0.18 & -0.55$\pm$0.10 & 1.00$\pm$0.06 & 1.18$\pm$0.21 \\
202821899 & 6024$\pm$125 & 3.91$\pm$0.20 & +0.30$\pm$0.10 & 1.42$\pm$0.20 & 2.10$\pm$0.60 \\
202843107 & 7493$\pm$65 & 3.69$\pm$0.15 & -0.19$\pm$0.06 & 1.93$\pm$0.16 & 3.29$\pm$0.63 \\  % hot star
%  202900527  FP Shporer+2017
203070421 & 6157$\pm$125 & 2.79$\pm$0.20 & -0.18$\pm$0.10 & 2.82$\pm$0.54 & 11.16$\pm$3.94 \\  %giant
203518244 & 6205$\pm$125 & 3.80$\pm$0.20 & -0.09$\pm$0.10 & 1.37$\pm$0.21 & 2.21$\pm$0.58 \\
203533312 & 6400$\pm$45 & 4.01$\pm$0.12 & -0.15$\pm$0.05 & 1.27$\pm$0.11 & 1.83$\pm$0.29 \\
203753577\tablenotemark{a} & 6171$\pm$47 & 4.01$\pm$0.11 & +0.05$\pm$0.05 & 1.29$\pm$0.11 & 1.84$\pm$0.28 \\
203771098 & 5644$\pm$40 & 4.35$\pm$0.11 & +0.50$\pm$0.05 & 1.03$\pm$0.02 & 1.12$\pm$0.12 \\ %metal rich
203826436 & 5379$\pm$125 & 4.55$\pm$0.20 & 0.00$\pm$0.10 & 0.87$\pm$0.05 & 0.84$\pm$0.06 \\
203867512 & 6367$\pm$47 & 3.86$\pm$0.12 & -0.15$\pm$0.05 & 1.39$\pm$0.14 & 2.23$\pm$0.43 \\
203929178 & 6820$\pm$64 & 4.34$\pm$0.12 & -0.62$\pm$0.05 & 1.09$\pm$0.03 & 1.20$\pm$0.14 \\
204221263 & 5643$\pm$40 & 4.30$\pm$0.11 & +0.34$\pm$0.05 & 1.03$\pm$0.02 & 1.18$\pm$0.14 \\
% 204914585  nans for radius. 
% 205029914  nans for radius.
205050711 & 7072$\pm$113 & 3.92$\pm$0.17 & -0.05$\pm$0.09 & 1.62$\pm$0.17 & 2.25$\pm$0.56 \\
205071984 & 5351$\pm$125 & 4.50$\pm$0.20 & +0.01$\pm$0.10 & 0.88$\pm$0.05 & 0.86$\pm$0.07 \\
205111664 & 5577$\pm$125 & 4.14$\pm$0.20 & -0.19$\pm$0.10 & 0.91$\pm$0.08 & 1.38$\pm$0.47 \\
205570849 & 5950$\pm$125 & 4.27$\pm$0.20 & -0.14$\pm$0.10 & 0.98$\pm$0.07 & 1.18$\pm$0.24 \\
205924614 & 4310$\pm$200 & 4.61$\pm$0.20 & -0.11$\pm$0.26 & 0.65$\pm$0.06 & 0.63$\pm$0.05 \\
205944181 & 5250$\pm$125 & 4.48$\pm$0.20 & +0.05$\pm$0.10 & 0.86$\pm$0.04 & 0.83$\pm$0.05 \\
205950854 & 5422$\pm$125 & 4.36$\pm$0.20 & -0.17$\pm$0.10 & 0.84$\pm$0.05 & 0.83$\pm$0.08 \\
205957328 & 5295$\pm$76 & 4.74$\pm$0.15 & +0.13$\pm$0.07 & 0.88$\pm$0.03 & 0.84$\pm$0.04 \\
% 205999468 & 5088$\pm$72 & 4.3$\pm$0.2 & -0.31$\pm$0.04 & 0.76$\pm$0.02 & 0.71$\pm$0.02 \\
% 206007892 & 5418$\pm$85 & 4.1$\pm$0.2 & +0.16$\pm$0.04 & 0.97$\pm$0.06 & 1.32$\pm$0.35 \\
% 206008091 & 5508$\pm$75 & 4.1$\pm$0.2 & -0.35$\pm$0.04 & 0.95$\pm$0.11 & 1.73$\pm$0.62 \\
206024342 & 5801$\pm$125 & 4.20$\pm$0.20 & -0.24$\pm$0.10 & 0.90$\pm$0.06 & 1.14$\pm$0.28 \\
206026136 & 4548$\pm$200 & 4.58$\pm$0.20 & -0.10$\pm$0.26 & 0.69$\pm$0.05 & 0.66$\pm$0.04 \\
% 206036749   nans for radius.
206038483 & 5597$\pm$77 & 4.12$\pm$0.16 & +0.26$\pm$0.08 & 1.02$\pm$0.06 & 1.35$\pm$0.27 \\
206049452 & 4447$\pm$200 & 4.59$\pm$0.20 & -0.27$\pm$0.26 & 0.65$\pm$0.05 & 0.63$\pm$0.05 \\
206055981 & 4544$\pm$200 & 4.58$\pm$0.20 & -0.30$\pm$0.26 & 0.65$\pm$0.06 & 0.63$\pm$0.05 \\
206082454 & 5573$\pm$125 & 4.79$\pm$0.20 & +0.11$\pm$0.10 & 0.95$\pm$0.05 & 0.93$\pm$0.07 \\
206096602 & 4561$\pm$200 & 4.58$\pm$0.20 & -0.18$\pm$0.26 & 0.69$\pm$0.06 & 0.66$\pm$0.05 \\
206103150 & 5392$\pm$125 & 4.08$\pm$0.20 & +0.34$\pm$0.10 & 1.00$\pm$0.08 & 1.39$\pm$0.37 \\
206114630 & 5097$\pm$44 & 4.53$\pm$0.11 & +0.06$\pm$0.05 & 0.82$\pm$0.02 & 0.78$\pm$0.02 \\
206125618 & 5351$\pm$125 & 4.37$\pm$0.20 & +0.04$\pm$0.10 & 0.88$\pm$0.05 & 0.87$\pm$0.08 \\
206135682 & 4838$\pm$42 & 4.74$\pm$0.20 & -0.12$\pm$0.10 & 0.73$\pm$0.03 & 0.70$\pm$0.03 \\
206245553 & 5819$\pm$42 & 4.35$\pm$0.11 & +0.08$\pm$0.05 & 1.02$\pm$0.03 & 1.12$\pm$0.13 \\
206311743 & 5146$\pm$125 & 4.04$\pm$0.20 & -0.42$\pm$0.10 & 0.93$\pm$0.09 & 2.38$\pm$0.42 \\ %metal poor with giant planet
206417197 & 5111$\pm$125 & 4.61$\pm$0.20 &  0.00$\pm$0.10 & 0.81$\pm$0.04 & 0.77$\pm$0.04 \\
%  FP Shporer+2017 206432863 & 5815$\pm$66 & 4.1$\pm$0.2 & +0.13$\pm$0.04 & 1.09$\pm$0.08 & 1.43$\pm$0.30 \\
\enddata
\tablenotetext{a}{Double-lined spectroscopic binary.}
\end{deluxetable}

%-------------------------------------------------------------------

%-------------------------------------------------------------------

%-------------------------------------------------------------------
\clearpage

%  updated radii and added new planets 28 Sep.
%  still need revised Sinc and Temps from Kane
%  V16 depths have no erorbars!
%  Use median value from C16 = 0.0025 in Rp/Rs

%\floattable
% \begin{longtable}
\begin{deluxetable*}{lllllrrrr}
\tabletypesize{\scriptsize}
\tablecolumns{9}
\tablewidth{0pt}
\tablecaption{Planet Candidate Properties}
\tablehead{
\colhead{EPIC ID} & \colhead{K2 ID} & \colhead{$P$ (d)} & \colhead{$a$ (au)} & \colhead{$R_p/R_*$} & \colhead{$R_p$ (\Rearth)} & \colhead{$S_{inc}$} & \colhead{$T_{eq}$ (K)} 
& \colhead{$T_{eq}$ (K) } \\
  &  &  &  &  &  & \colhead{F$_{\oplus}$} & hot dayside & well-mixed 
}
\startdata
\label{tab:planets}
201155177   & K2-42b  & 6.68796$\pm$0.00093 & 0.061 &  0.0304$\pm$0.0028\tablenotemark{a} & 2.19$\pm$0.25 & 47.229  & 868.5 & 730.3  \\
201291843   &         & 40.70206194        & 0.194  & 0.00536$\pm$0.00028\tablenotemark{b} & 0.34$\pm$0.03 & 2.556  & 418.9  & 352.2  \\
201393098   & K2-7b   & 28.6777$\pm$0.0086 & 0.182  & 0.0177$\pm$0.0018\tablenotemark{a} & 3.38$\pm$0.84 & 84.638 & 1004.8  & 845.0   \\
201403446   & K2-46b  & 19.1541$\pm$0.0041 & 0.140 &  0.0145$\pm$0.001\tablenotemark{a} & 2.18$\pm$0.47 & 125.074  & 1107.9 &  931.6  \\
201407812\tablenotemark{e}   &         & 2.8268121          & 0.037 &  0.4560\tablenotemark{c} & 84.68$\pm$12.00 & 2419.735 & 2323.5  & 1953.8  \\
201516974   &         & 36.762337          & 0.221 &  0.03712\tablenotemark{c} & 33.21$\pm$6.84 &  719.553 & 1715.8  & 1442.8  \\
201546283   & K2-27b  & 6.771315$\pm$0.000079 & 0.068 &  0.0474$\pm$0.00093\tablenotemark{a} & 4.53$\pm$0.24 & 115.165  & 1085.3 & 912.6 \\
201606542   &         & 0.444372$\pm$0.000042 & 0.011 & 0.0136$\pm$0.002\tablenotemark{d} & 1.33$\pm$0.21 & 4796.510 & 2757.0 &  2318.3  \\
% 201754305.1 & K2-16b  & 7.61856$\pm$0.00096 & 0.068 & 0.0268$\pm$0.0022\tablenotemark{a} & 1.95$\pm$0.18 & 41.263  &  839.6 &  706.1 \\
% 201754305.2 & K2-16c  & 19.074540           & 0.125 & 0.03269\tablenotemark{c} & 2.37$\pm$0.21 & 12.137  & 618.3 &  520.0  \\
201855371   & K2-17b  & 17.9654$\pm$0.0017 & 0.116 & 0.029$\pm$0.0025\tablenotemark{a} & 2.00$\pm$0.22 & 9.692  & 584.5 &  491.5 \\ 
201912552   & K2-18b  & 32.9418$\pm$0.0021 & 0.161 & 0.0517$\pm$0.0021\tablenotemark{a} & 2.82$\pm$0.32 & 2.453  & 414.6 &  348.6  \\
202634963\tablenotemark{e}   &         & 28.707623          & 0.181 & 0.2136\tablenotemark{c} & 27.05$\pm$5.14 & 61.455  & 927.6 &  780.0 \\
202675839   &         & 15.4715$\pm$0.0036 & 0.127 & 0.044$\pm$0.075\tablenotemark{a} & 7.35$\pm$12.64  & 144.466 & 1148.5  & 965.8  \\
202688980\tablenotemark{e}   &         & 1.45566370         & 0.025 & 0.02958$\pm$0.00036\tablenotemark{b}  & 3.82$\pm$0.67 & 3438.517 & 2536.9  & 2133.2  \\
202821899   &         & 4.4743465          & 0.060 & 0.03229\tablenotemark{c} & 7.40$\pm$2.18 & 1449.133 & 2044.0 &  1718.8 \\
202843107   &         & 2.1989041          & 0.041 & 0.6032\tablenotemark{c} & 216.57$\pm$41.66 & 17909.266  & 3832.4 & 3222.7 \\
203070421   &         & 1.7359447          & 0.040 & 0.02551\tablenotemark{c} & 31.08$\pm$11.40 & 100522.547  & 5898.9  & 4960.3 \\
203518244   &         & 0.84112570         & 0.019 & 0.01098\tablenotemark{c} & 2.65$\pm$0.92 & 17425.977  & 3806.3  & 3200.7 \\
203533312   &         & 0.17566$\pm$0.000183 & 0.007  & 0.0248$\pm$0.001\tablenotemark{d} & 4.95$\pm$0.82 & 113835.766  & 6085.2  & 5117.0  \\
203753577\tablenotemark{e}   &         & 3.4007758          & 0.050 & 0.06863\tablenotemark{c} & 13.78$\pm$2.16 & 1886.993  & 2183.5  & 1836.1  \\
203771098.1 & K2-24c  & 42.36301$\pm$0.00072 & 0.241 & 0.05913$\pm$0.00053\tablenotemark{a} & 7.25$\pm$0.75 & 19.848 & 699.3  & 588.0  \\
203771098.2 & K2-24b  & 20.88526$\pm$0.00042 & 0.150 & 0.04264$\pm$0.00081\tablenotemark{a} & 5.23$\pm$0.55 & 50.962 & 885.1  & 744.3  \\
203826436.1 & K2-37b  & 4.44118$\pm$0.00074 & 0.050 & 0.0174$\pm$0.0015\tablenotemark{a} & 1.60$\pm$0.18 & 210.138  & 1261.3  & 1060.7  \\
203826436.2 & K2-37c  & 6.42973$\pm$0.00043 & 0.065 & 0.0276$\pm$0.0018\tablenotemark{a} & 2.54$\pm$0.25 & 128.305 & 1115.0  & 937.6 \\
203826436.3 & K2-37d  & 14.0919$\pm$0.0015 & 0.109 & 0.0271$\pm$0.0021\tablenotemark{a} & 2.50$\pm$0.27 & 45.069  & 858.4  & 721.8  \\
203867512   &         & 28.465633          & 0.203  & 0.1642\tablenotemark{c} & 39.98$\pm$7.70 & 175.975 & 1206.6  & 1014.6  \\
203929178   &         & 1.153886$\pm$0.000028 & 0.022  & 0.53$\pm$0.23\tablenotemark{a} & 69.23$\pm$31.17 & 5693.021 & 2877.7  & 2419.8  \\
204221263.1 & K2-38b  & 4.01628$\pm$0.00044 & 0.050 & 0.01329$\pm$0.00099\tablenotemark{a} & 1.72$\pm$0.24 & 508.436  & 1573.1  & 1322.8  \\ 
204221263.2 & K2-38c  & 10.56098$\pm$0.00081 & 0.095 & 0.0195$\pm$0.0014\tablenotemark{a} & 2.52$\pm$0.35 & 140.087  & 1139.7  & 958.4 \\
% 204914585   &         & 18.357773            & 0.135 & 0.01924\tablenotemark{c} & 2.04$\pm$0.36 &   & 863.5  &  726.1  \\
205050711   &         & 4.30221683             & 0.061  & 0.02613$\pm$0.00074\tablenotemark{b}  & 6.41$\pm$1.62 & 3029.473 & 2457.8  & 2066.8  \\
205071984.1 & K2-32b  & 8.99213$\pm$0.00015 & 0.081 & 0.0556$\pm$0.0014\tablenotemark{a} & 5.24$\pm$0.43 & 86.520  & 1010.4 &  849.6  \\
205071984.2 & K2-32c  & 20.6602$\pm$0.0016 & 0.141 & 0.0326$\pm$0.0021\tablenotemark{a} & 3.07$\pm$0.31 & 28.538  & 765.7  &  643.9  \\
205071984.3 & K2-32d  & 31.7154$\pm$0.0020 & 0.188 & 0.0371$\pm$0.0031\tablenotemark{a} & 3.50$\pm$0.40 & 16.115  & 663.8  &  558.2  \\
205111664   &         & 15.937378          & 0.120 & 0.02135\tablenotemark{c} & 3.22$\pm$1.16 & 125.627  & 1109.1 & 932.6  \\
205570849   &         & 16.8580$\pm$0.0011 & 0.128 & 0.047$\pm$0.057\tablenotemark{a} & 6.04$\pm$7.43 & 96.813  & 1039.2 & 873.8 \\
205924614   & K2-55b  & 2.849258$\pm$0.000033 & 0.034 & 0.0552$\pm$0.0013\tablenotemark{a} & 3.81$\pm$0.30 & 107.097  & 1065.7 & 896.2 \\
205944181   &         & 2.475527$\pm$0.000083 & 0.034 & 0.38$\pm$0.35\tablenotemark{a} & 34.55$\pm$31.89 & 412.831  & 1493.3 & 1255.7  \\
205950854   &         & 15.854120            & 0.116 & 0.02208\tablenotemark{c} & 2.00$\pm$0.29 & 41.132  & 839.0 &  705.5  \\
205957328   &         & 14.353347            & 0.111 & 0.02383\tablenotemark{c} & 2.19$\pm$0.25 & 40.436  & 835.4  & 702.5  \\
206024342.1 &         & 14.6370$\pm$0.0021   & 0.113 & 0.0249$\pm$0.0015\tablenotemark{a} & 3.09$\pm$0.79 & 107.512 & 1066.8  & 897.0  \\
206024342.2 &         & 0.91165670           & 0.018 & 0.01593\tablenotemark{c} & 1.98$\pm$0.58 & 4354.620 & 2691.2  & 2263.0  \\
% 205999468   &         & 12.2631$\pm$0.0011 & 0.095 & 0.0238$\pm$0.0022\tablenotemark{a} & 1.86$\pm$0.18 & 32.892    & 793.4 &  667.1  \\
% 206007892   &         & 6.3796100           & 0.067 & 0.01388\tablenotemark{c} & 2.00$\pm$0.65 & 319.415  & 1400.5 &  1177.7  \\
% 206008091.1 &         & 7.5767206          & 0.074 & 0.01287\tablenotemark{c} & 2.43$\pm$0.99 & 460.018   & 1534.3  & 1290.1 \\
% 206008091.2 &         & 12.400732          & 0.103 & 0.01629\tablenotemark{c} & 3.08$\pm$1.20 & 238.499   & 1301.9  & 1094.8  \\
206026136   & K2-57b  & 9.0063$\pm$0.0013 & 0.075 & 0.0308$\pm$0.0028\tablenotemark{a} & 2.23$\pm$0.25 & 30.411  & 778.0 & 654.2  \\ 
206038483   & K2-60b  & 3.002627$\pm$0.000018 & 0.041 & 0.06191$\pm$0.00035\tablenotemark{a} & 9.12$\pm$1.80 & 959.652  & 1843.9 & 1550.5  \\
206049452   &         & 14.454495             & 0.101 & 0.02923\tablenotemark{c} & 2.01$\pm$0.23 & 14.350 & 644.8  & 542.2  \\
206055981   &         & 20.643928            & 0.128 & 0.03129\tablenotemark{c} & 2.16$\pm$0.24 & 9.585  & 582.9 & 490.2  \\
206082454.1 &         & 14.317001            & 0.113 & 0.01714\tablenotemark{c} & 1.73$\pm$0.29 & 57.351 & 911.7  & 766.6  \\
206082454.2 &         & 29.626402            & 0.184 & 0.03282\tablenotemark{c} & 3.32$\pm$0.36 & 21.749  & 715.4 & 601.6 \\
206096602.1 & K2-62b  & 6.67202$\pm$0.00028 & 0.061 & 0.0271$\pm$0.0017\tablenotemark{a} & 1.95$\pm$0.18 & 47.468  & 869.6 & 731.2  \\
206096602.2 & K2-62c  & 16.1966$\pm$0.0012 & 0.111 & 0.0269$\pm$0.0019\tablenotemark{a} & 1.94$\pm$0.19 & 14.549  & 647.0  & 544.1  \\
206103150.1 & WASP-47b & 4.159221$\pm$0.000015 & 0.051  & 0.10214$\pm$0.0003\tablenotemark{a} & 15.54$\pm$4.08 & 627.497 & 1658.1  & 1394.3  \\ 
206103150.2 & WASP-47d & 9.03164$\pm$0.00064 & 0.085 & 0.026$\pm$0.0015\tablenotemark{a} & 3.96$\pm$1.06 & 223.155  & 1280.4  & 1076.7  \\
206103150.3 & WASP-47e & 0.789518$\pm$0.000060 & 0.017  & 0.01344$\pm$0.00088\tablenotemark{a} & 2.04$\pm$0.55 & 5751.909  & 2885.1  & 2426.0  \\
206114630   &         & 7.4448754           & 0.070 & 2.65$\pm$0.23\tablenotemark{c} & 2.65$\pm$0.23 &  75.689  & 977.2  & 821.7  \\
206125618   & K2-64b  & 6.53044$\pm$0.00067 & 0.065 & 0.0259$\pm$0.0017\tablenotemark{a} & 2.47$\pm$0.27 & 132.337 & 1123.6 & 944.9  \\ 
206135682   &         & 5.0258310           & 0.052 & 0.01961\tablenotemark{c} & 1.49$\pm$0.20 & 91.188  & 1023.7 & 860.9  \\
206245553   & K2-73b  & 7.49543$\pm$0.00059 & 0.075 & 0.021$\pm$0.0012\tablenotemark{a} & 2.56$\pm$0.33 & 223.737  & 1281.3 & 1077.4  \\ 
206311743   &         & 4.31444335          & 0.051 & 0.03877$\pm$0.00040\tablenotemark{b}  & 10.06$\pm$1.79 & 1404.529  & 2028.1  & 1705.4  \\
206417197   &         & 0.442094$\pm$0.000086 & 0.011 & 0.0138$\pm$0.001\tablenotemark{d} & 1.15$\pm$0.10 & 3122.265  & 2476.4 & 2082.4  \\
% 206432863   & K2-76b  & 23.9726$\pm$0.0011 & 0.168 & 0.0793$\pm$0.0017\tablenotemark{a} & 12.42$\pm$2.65 & 74.020  & 971.7  & 817.1 \\
\enddata
\tablenotetext{a}{\citet{c16}}
\tablenotetext{b}{\citet{barros16}}
\tablenotetext{c}{\citet{v16}} 
\tablenotetext{d}{\citet{adams16}} 
\tablenotetext{e}{Double-lined spectroscopic binary star.}
\end{deluxetable*}
% \end{longtable}  this is bullshit

%-------------------------------------------------------------------
\clearpage
%  updated plots 28 Sep

%-------------------------------------------------------------------

%-------------------------------------------------------------------
%  updated this plot 28 Sep

%-------------------------------------------------------------------

%-------------------------------------------------------------------

%\includegraphics[angle=0, scale=0.5]{fluxplot.pdf}
%\caption{Insolation (relative to earth) as a function of host-star 
%effective temperature for the 57 planet candidates described in 
%Table~\ref{tab:planets}. }
%\label{fig:flux}
%\end{figure}

\clearpage
%-------------------------------------------------------------------

\begin{deluxetable}{lrrrr}
\tabletypesize{\scriptsize}
\tablecolumns{5}
\tablewidth{0pt}
\tablecaption{Habitable Zone Boundaries for Planet Candidate Host Stars}
\tablehead{
\colhead{EPIC ID} & \colhead{Inner HZ (au) } & \colhead{Inner HZ (au) } & \colhead{Outer HZ (au) } & \colhead{Outer HZ (au) }\\ 
   & \colhead{optimistic} & \colhead{conservative} & \colhead{optimistic} & \colhead{conservative}
 }
\startdata
\label{habzones}
201155177 &  0.33 & 0.42 & 0.78 & 0.82 \\
201291843 &  0.25 & 0.32 & 0.60 & 0.63 \\
201393098 &  1.27 & 1.60 & 2.84 & 2.99 \\
201403446 &  1.13 & 1.43 & 2.50 & 2.64 \\
201407812 &  1.34 & 1.70 & 2.98 & 3.14 \\
201516974 &  4.67 & 5.91 & 10.74 & 11.33 \\
201546283 &  0.56 & 0.71 & 1.27 & 1.34 \\
201606542 &  0.59 & 0.75 & 1.34 & 1.41 \\
201855371 &  0.29 & 0.37 & 0.68 & 0.72 \\
201912552 &  0.20 & 0.26 & 0.48 & 0.51 \\
202634963 &  1.03 & 1.30 & 2.26 & 2.39 \\
202675839 &  1.15 & 1.45 & 2.56 & 2.70 \\
202688980 &  1.06 & 1.34 & 2.34 & 2.46 \\
202821899 &  1.68 & 2.13 & 3.73 & 3.93 \\
202843107 &  3.78 & 4.79 & 8.26 & 8.71 \\
203070421 &  9.29 & 11.76 & 20.56 & 21.69 \\
203518244 &  1.87 & 2.37 & 4.13 & 4.36 \\
203533312 &  1.62 & 2.05 & 3.57 & 3.76 \\
203753577 &  1.53 & 1.94 & 3.40 & 3.58 \\
203771098 &  0.81 & 1.03 & 1.82 & 1.92 \\
203826436 &  0.56 & 0.71 & 1.27 & 1.34 \\
203867512 &  1.95 & 2.47 & 4.31 & 4.54 \\
203929178 &  1.18 & 1.49 & 2.59 & 2.73 \\
204221263 &  0.85 & 1.08 & 1.91 & 2.01 \\
205050711 &  2.33 & 2.95 & 5.10 & 5.38 \\
205071984 &  0.58 & 0.73 & 1.31 & 1.38 \\
205111664 &  1.02 & 1.30 & 2.30 & 2.42 \\
205570849 &  0.93 & 1.18 & 2.08 & 2.19 \\
205924614 &  0.28 & 0.36 & 0.67 & 0.71 \\
205944181 &  0.53 & 0.68 & 1.21 & 1.28 \\ 
205950854 &  0.57 & 0.72 & 1.29 & 1.36 \\
205957328 &  0.54 & 0.69 & 1.23 & 1.30 \\
206024342 &  0.88 & 1.11 & 1.96 & 2.07 \\
206026136 &  0.33 & 0.42 & 0.77 & 0.81 \\
206038483 &  0.96 & 1.22 & 2.16 & 2.28 \\
206049452 &  0.31 & 0.39 & 0.72 & 0.76 \\
206055981 &  0.32 & 0.40 & 0.74 & 0.78 \\
206082454 &  0.65 & 0.83 & 1.47 & 1.55 \\
206096602 &  0.34 & 0.43 & 0.79 & 0.83 \\
206103150 &  0.97 & 1.23 & 2.20 & 2.32 \\
206114630 &  0.47 & 0.60 & 1.08 & 1.14 \\
206125618 &  0.58 & 0.73 & 1.31 & 1.38 \\
206135682 &  0.39 & 0.49 & 0.90 & 0.95 \\
206245553 &  0.84 & 1.07 & 1.88 & 1.99 \\
206311743 &  1.48 & 1.87 & 3.37 & 3.55 \\
206417197 &  0.46 & 0.58 & 1.05 & 1.11 \\
\enddata
\end{deluxetable}

\end{document}